\documentclass{aa}  

\usepackage{graphicx}
\usepackage{txfonts}
\usepackage{placeins}
\newcommand{\asec}{$^{\prime\prime}$}

\def\SigmaH2{$\Sigma $(${\rm H_2}$)}
\def\r1415{$^{14}$N/$^{15}$N}

\def\cyclic{{\it c-}C$_3$H$_2$}
\def\cyclicII{{\it c-}C$_3$H}
\def\cIIh{C$_2$H}
\def\cIVh{C$_4$H}
\def\CIII{HC$_3$N}
\def\CV{HC$_5$N}
\def\methyl{CH$_3$CCH}
\def\cyanide{CH$_3$CN}
\def\H{N$_{2}$H$^{+}$}

\def\15N{$^{15}$NNH$^+$}
\def\N15{N$^{15}$NH$^+$}

\def\AMM{NH$_3$}

\def\METH{CH$_3$OH}

\def\H13CN{\mbox{H$^{13}$CN}}
\def\HII{H{\sc ii}}

\def\kms{\mbox{km~s$^{-1}$}}
\def\cmc{cm$^{-3}$}
\def\cmq{cm$^{-2}$}

\def\solm{\mbox{M$_\odot$}}

\def\Ntot{$N_{\rm tot}$}
\def\Tex{\mbox{$T_{\rm ex}$}}

\def\Tk{\mbox{$T_{\rm k}$}}

\def\kms{km\,s$^{-1}$}

\def\HMSC{G034}
\def\HMPO{AFGL}
\def\UCHII{G589}

\begin{document}

   \title{CHemical Evolution in MassIve star-forming COres (CHEMICO)}

   \subtitle{I. Evolution of the temperature structure}

   \author{F. Fontani
          \inst{1,2,3}
        \and
        V.M. Rivilla\inst{4}
        \and
        E. Roueff\inst{2}   
        \and
        H. Mart\'in-Caballero\inst{4}
        \and
        L. Bizzocchi\inst{5}
        \and
        L. Colzi\inst{4}
        \and
        \'A. Lopez-Gallifa\inst{4}
        \and
        M.T. Beltr\'an\inst{1}
        \and
        P. Caselli\inst{3}
        \and
        C. Mininni\inst{6}
        \and
        A. Vasyunin\inst{7}
          }

    \institute{INAF - Osservatorio Astrofisico di Arcetri,
    Largo E. Fermi 5,
    I-50125, Florence (Italy)\\
              \email{francesco.fontani@inaf.it}
         \and
         LUX, Observatoire de Paris, PSL Research University, CNRS,Sorbonne
         Universit\'e,F-92190 Meudon (France)
         \and
         Max-Planck-Institut f\"{u}r extraterrestrische Physik, Giessenbachstra\ss e 1, 85748 Garching bei M\"{u}nchen, Germany
         \and
        Centro de Astrobiolog\'ia (CAB), CSIC-INTA, Ctra. de Ajalvir Km. 4, 28850, Torrej\'on de Ardoz, Madrid, Spain
        \and
        Dipartimento di Chimica “Giacomo Ciamician”, Universit\'a di Bologna, Via P. Gobetti 85, I-40129 Bologna (Italy) 
        \and
        INAF - Istituto di Astrofisica e Planetologia Spaziali, Via Fosso del Cavaliere 100, I-00133 Roma (Italy)
        \and
        Research Laboratory for Astrochemistry, Ural Federal University, Kuibysheva St. 48, Yekaterinburg 620026, Russia
             }

   \date{Received xxx; accepted xxx}

 
  \abstract
   {Increasing evidence shows that most stars in the Milky Way, including the Sun, are born in star-forming
regions containing also high-mass stars, but due to both observational and theoretical challenges, our comprehension of their
chemical evolution is far less clear than that of their low-mass counterparts.}
   {We present here the project "CHemical Evolution of MassIve star-froming COres" (CHEMICO).
   The project aims at investigating any aspect
of the chemical evolution of high-mass star-forming cores by observing representatives of the three main evolutionary categories: 
high-mass starless cores, high-mass protostellar objects, and ultra-compact \HII\ (UCHII) regions.}
   {We carried out an unbiased spectral line survey of the entire bandwidth at 3, 2, and 1.2~mm with the 30m telescope of the Insitut de Radioastronomie millim\'etrique towards three targets that represent the three evolutionary stages.} 
   {The number of lines and species detected increases with evolution.
   In this first work, we derive the temperature structure of the targets through the analysis of the carbon-bearing species \cIIh, \cyclicII, \cyclic, \cIVh, \methyl, \CIII,  \cyanide\ and \CV. 
   The excitation temperature, \Tex, increases with evolution in each species, although not in the same way. 
   Hydrocarbons and carbon-chains tend to be associated with the smallest \Tex\ values and enhancements with evolution, while
   cyanides are associated with the highest \Tex\ values and enhancements.
   In each target, the higher the number of atoms in the molecule, the higher \Tex\ tends to be.}
   {The temperature structure evolves from a cold ($\sim 20$~K), uniform envelope traced by simple hydrocarbons in the high-mass starless core stage, to a more stratified envelope in the protostellar stage made by a hot core ($\geq 100$~K), an intermediate shell with \Tex\ $\sim30-60$~K and a larger cold envelope, to finally a hot core surrounded only by a cold envelope in the UCHII stage.
   This suggests a steepening of the \Tex\ radial profile as a function of time.}

\keywords{astrochemistry – line:
          identification – 
          ISM: molecules – 
          stars: formation   
   }

   \maketitle
%

\section{Introduction}
\label{intro}
Nowadays growing evidence indicates that our Sun was born in a crowded
stellar cluster including high-mass (HM) stars, i.e. stars more massive than $\sim 8$~\solm\ (e.g. Adams~\citeyear{adams10},
Pfalzner et al.~\citeyear{pfalzner15}, Jensen et al.~\citeyear{jensen19}, Arzoumanian et al.~\citeyear{arzoumanian23}).
The case of the Sun is not peculiar, 
because it is now well known that most stars are born in rich clusters (e.g.~Carpenter~\citeyear{carpenter00}; 
Lada \& Lada~\citeyear{lel03}) which likely included HM stars (e.g.~Rivilla et al.~\citeyear{rivilla14}).
Therefore, the study of the properties of HM star-forming regions can give us important information not only on the formation of HM stars but also of the 
heritage of both the Solar System and most stars in the Milky Way.

   
\begin{figure}
   \centering
   \includegraphics[width=9cm]{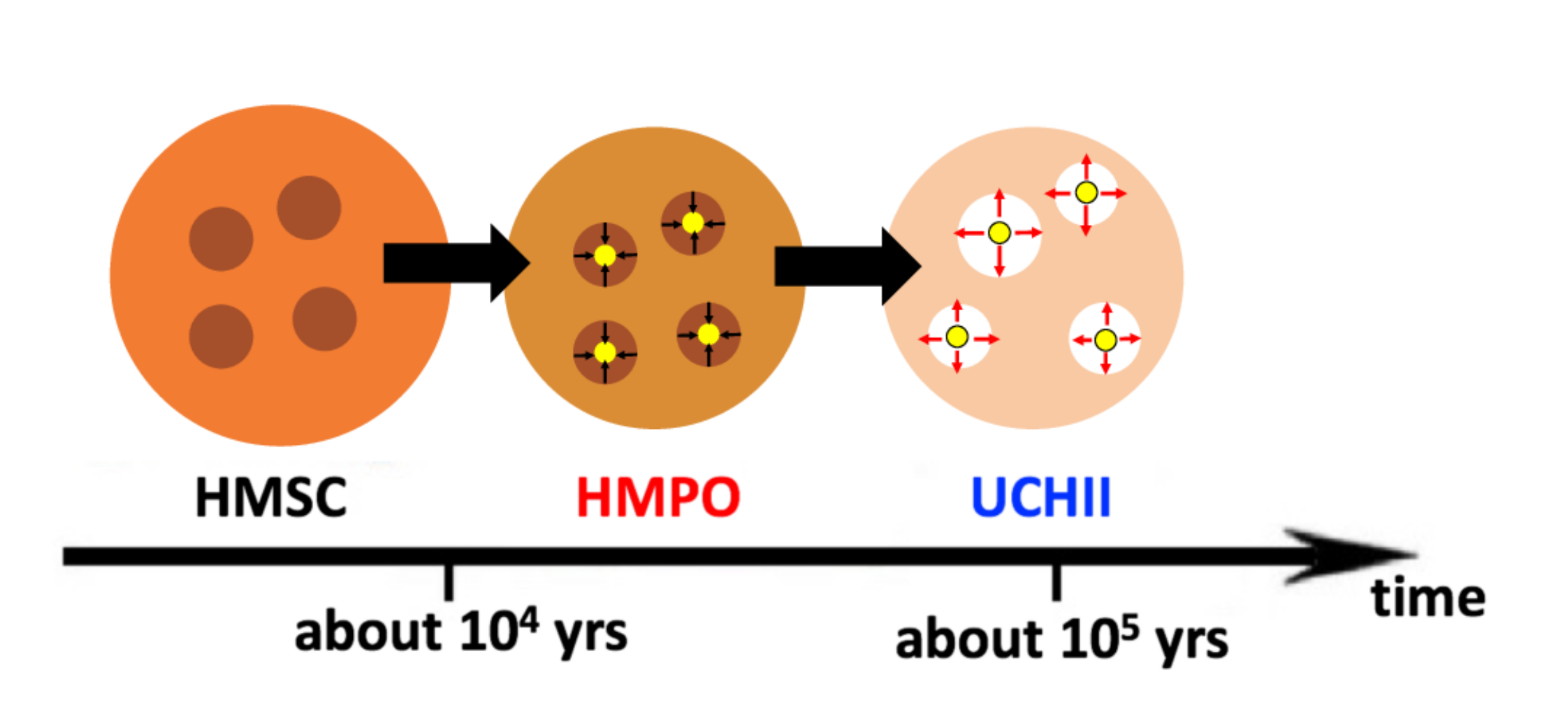}
      \caption{Scheme of the coarse evolutionary classification for HM star-forming cores in HMSCs,
HMPOs, and UCHIIs \citep[][adapted from Motte et al. 2018]{colziPhD}.
              }
    \label{fig:sequence}
\end{figure}

Our knowledge of how HM stars 
are born is still limited, owing to both observational and theoretical problems 
(e.g.~Zinnecker \& Yorke~\citeyear{zey07}; Tan et al.~\citeyear{tan14}; Krumholz~\citeyear{krumholz15}; Motte et al.~\citeyear{motte18}). 
Observations are challenging because HM star-forming cores, i.e. molecular compact structures, which have the potential to form HM stars and/or 
clusters, are fewer with respect to their low-mass counterparts, smaller in angular dimensions (being located at distances $\geq 1$~kpc), and typically 
"polluted" by large amounts of ambient gas and/or nearby star formation activity. Theoretical problems arise mostly from the fact that HM star-forming cores 
evolve in timescales of $\leq 10^5$ years, typically shorter than those of their low-mass counterparts, and in particular accretion lasts longer than contraction \citep{yorke86}.
This implies that HM stars switch on while still accreting, and the radiation pressure of the embryo star 
could be sufficient to impede any further accretion \citep{les71,aet74}. Even though models can solve this problem \citep[e.g.][]{met03,bonnell04}, the steps of the HM star-formation process are still debated and not clearly separated into evolutionary classes
as in the low-mass star formation process (see,  e.g.,  Shu,  Adams  \&  Lizano~\citeyear{shu87};  Andr\'e,  Ward-Thompson \& Barsony~\citeyear{andre00}). 
Therefore, finding a chemical evolutionary sequence would be extremely useful in defining a physical evolutionary sequence still uncertain.

Several attempts to empirically classify pc-scale HM star-forming cores in different evolutionary stages have been 
proposed (e.g.~Tan et al.~\citeyear{tan14}; Motte et al.~\citeyear{motte18}). Following the sequence of low-mass star-forming cores, 
they all summarily divide the evolution into three main phases (see Fig.~\ref{fig:sequence}): 
(1) High-Mass Starless Cores (HMSCs):
these objects, mostly found in infrared-dark, dense molecular clouds,  
are in a physical stage immediately before (or at the very beginning of) the gravitational collapse. 
They are characterised by low temperatures (\Tk $\sim 10-20$~K), high densities
($n\geq 10^{3}-10^{5}$~\cmc, depending on the size), and show evidence of outflows or signs of star formation at very early stage, or absent;
In these early cold phases, atoms and
simple molecules are thought to freeze-out on dust grain surfaces, hence surface 
chemistry is very active and gas-phase
chemistry is inhibited, particularly 
neutral-neutral and endothermic reactions;
(2) High-Mass Protostellar Objects (HMPOs): these are collapsing cores with evidence of deeply embedded protostar(s),
characterised typically by densities and temperatures higher than in the previous stage
($n\simeq 10^6$~\cmc, \Tk $\geq 20$~K).
In this warm(er) environment, the molecules in the mantles 
sublimate back in the gas-phase, and reactions not efficient at low temperatures start to proceed,
and form new complex molecules. Also, collimated jets and molecular outflows from 
the protostar(s) can trigger local high-energy and (likely) non-equilibrium chemistry typical of shocked gas.
(3) Ultra-compact \HII\ regions (UCHIIs): 
these cores contain at least one embedded Zero-Age-Main-Sequence star associated with an expanding \HII\ region, whose 
surrounding molecular cocoon ($n\geq 10^5$~\cmc, 
\Tk $\sim 20 - 100$~K) can be affected by its progressive expansion and by heating and irradiation from the central star.

Chemical and physical evolution are expected to proceed in parallel, but only an unbiased survey of the whole (sub-)mm band towards sources in a well-established evolutionary stage 
can fully characterise the chemical evolution from one stage to another. In particular, identifying evolutionary diagnostic tools to be used in future 
studies of larger samples would be crucial. 
These studies can also provide constraints on chemical-dynamical models (e.g. Sipil\"{a} \& Caselli~\citeyear{sec18})
of HM star-forming regions, thus elucidating 
important chemical processes occurring at the three different evolutionary stages, which in turn will help us to link such stages in a unified scenario.
Previous studies have highlighted potential evolutionary indicators, particularly useful to trace specific phases.
For example, some deuterated molecules are excellent tracers of the earliest stages \citep[e.g.][]{fontani11,fontani15a,sabatini24}, while some complex organic molecules (COMs) or sulphur-bearing species appear to trace later phases \citep[e.g.][]{coletta20,fontani23}.

We present the project "CHemical Evolution in Massive star-forming COres (CHEMICO)", in which we perform the first full spectral survey in the millimetre bands of the Institut de RadioAstronomie Millim\'etrique (IRAM) 30m telescope
towards three dense cores that best represent the three evolutionary stages of HM star-formation. 
The observed lines will allow us to derive the column densities of all molecules detected from all observable transitions. 
From the molecular column densities, the 
abundances will also be derived from the H$_2$ column densities, $N$(H$_2$), already calculated for all 
objects (Fontani et al.~\citeyear{fontani18}). 
Because chemical evolution is far more studied for low-mass star-forming cores \citep[see, e.g. the reviews by][]{cec12,jorgensen20}, this project
aims also to compare the chemical evolution in the low- and high-mass regime.

In this paper, we present the data, the sample, and the method to identify and analyse the molecules.
We also present the analysis of some carbon-chain and hydrocarbon species.
Carbon-chains are ubiquitous in the interstellar medium and claimed to be linked to the evolutionary stage in low-mass star-forming regions (see~e.g. Taniguchi et al.~\citeyear{taniguchi24} and references therein).
The fact that they can be formed from atomic carbon \citep[e.g.][]{suzuki92,taniguchi19} suggests that their formation is favoured before CO formation, namely in early stages of star formation, in agreement with
the detection of long hydrocarbons and cyanopolyynes in cold dark cores \citep{loren84,cernicharo21,loomis21}.
However, this simple scenario is challenged by the fact that carbon-chains and carbon-rich species (that we define in this study as molecules with at least 3 carbon atoms and no oxygen) are abundant also in warm ($T\geq 30$~K) low-mass and high-mass protostellar cores \citep[e.g.][]{sey13,taniguchi23}.
Reactions starting, for example, from gaseous methane (CH$_4$) evaporated from ice mantles were invoked to explain their abundances in evolved protostellar stages \citep[e.g.][]{sey13}.
Carbon chains increase their abundances also in presence of locally accelerated cosmic rays \citep[e.g.][]{fontani17} as well in the cavity walls of outflows \citep[e.g.][]{zhang18,tychoniec21}.
Furthermore, carbon-chain and other carbon-rich species such as \methyl\ and \CIII, are commonly used as temperature tracers
\citep[e.g.][]{fontani02,taniguchi19} in star-forming cores, thanks to their molecular structure and large number of rotational transitions observable in the (sub-)millimetre regime.
Understanding the temperature variation in the targets is critical to properly modelling the chemistry.
Therefore, in this first paper we analyse carbon-chains and carbon-rich species not only to inspect how their abundances vary with evolution, but also to find variations in the thermal structure of the cores.
In this respect, we include in this study the analysis of \cyanide, a carbon-bearing COM that is neither a carbon-chain nor a carbon-rich species, but it is a good gas thermometer and a species sensitive to evolution \citep[e.g.][]{mininni21}.

The paper is organised as follows:
Sect.~\ref{sample} presents the source sample.
In Sect.~\ref{obsandredu} we describe the observations and the procedures adopted for data reduction and line identification and analysis. 
Sect.~\ref{res} shows the line and physical parameters derived from the detected carbon-bearing species. 
In Sect.~\ref{discu} we discuss our results focusing the attention on the evolution of both the temperature structure(s) and the molecular abundances. 
Conclusions and future steps of this project will be illustrated in Sect.~\ref{conc}.


\section{Sample}
\label{sample}

The sources are part of a sample extensively studied in the past decade, consisting of 27 carefully
selected targets \citep{fontani11}, and divided into the three evolutionary classes described in Sect.~\ref{intro}: 11 HMSCs, 9 HMPOs, and 7 UCHIIs.
We have investigated so far specific aspects of the chemical evolution, such as
D and N fractionation (e.g. Fontani et al.~\citeyear{fontani15a},~\citeyear{fontani15b}, Colzi et al.~\citeyear{colzi18a},~\citeyear{colzi18b}, Rivilla et al.~\citeyear{rivilla20a}), P-bearing molecules (Fontani et al.~\citeyear{fontani16},~\citeyear{fontani19}, Mininni et al.~\citeyear{mininni18}), the rare ions HOCO$^+$ and HCNH$^+$ (Fontani et al.~\citeyear{fontani18},~\citeyear{fontani21}), and some COMs (dimethyl ether, methyl formate and formamide, Coletta et al.~\citeyear{coletta20}).

\begin{table*}
\caption[]{Source coordinates and basic physical parameters.}
    \label{tab:sources}
         \begin{tabular}{lcccccc}
            \hline
            \noalign{\smallskip}
            Source      &  R.A.\tablefootmark{a} & Dec.\tablefootmark{a} & V$_{\rm LSR}$\tablefootmark{b} & $d$\tablefootmark{b} & $L$\tablefootmark{b} & $N({\rm H_2})$\tablefootmark{c} \\
                &  ($h:m:s$) & ($\circ:\prime:\prime\prime$) & (\kms) & (kpc) & (L$_{\odot}$) & (\cmq) \\
            \noalign{\smallskip}
            \hline
            \noalign{\smallskip}
            HMSC & & & & & & \\
            G034G2(MM2) [G034] & 18:56:50.0 & 01:23:08  &  +43.6   & 2.9 & $10^{1.6}$ & 4.1$\times 10^{22}$ \\
            \noalign{\smallskip}
            HMPO & & & & & & \\
           AFGL5142--MM [AFGL] & 05:30:48.0 & 33:47:54 &  $-3.9$ & 1.8 & $10^{3.6}$ & 1.0$\times 10^{23}$ \\
            \noalign{\smallskip}
            UCHII & & & & & & \\
            G5.89--0.39 [G589] & 18:00:30.5 & $-24$:04:01 & +9.0 & 1.3 & $10^{5.1}$ & 5.5$\times 10^{23}$ \\
            \noalign{\smallskip}
            \hline
         \end{tabular}
    \tablefoot{\tablefoottext{a}{J2000;}
    \tablefoottext{b}{Derived from either observations of $^{13}$CO (G034) or \AMM\ inversion transitions (AFGL, G589). Reference works in \citet{fontani11};}
    \tablefoottext{c}{From \citet{fontani18}, computed from the dust thermal continuum emission images at 850~$\mu$m obtained with SCUBA at the James Clerk Maxwell Telescope (JCMT) by \citet{difrancesco08}.}
    }
   \end{table*}

We selected G034G2(MM2) (hereafter G034) as representative
of HMSC, AFGL5142--MM (herefafter AFGL) as HMPO, and G5.89-0.39 (hereafter G589) as UCHII by applying these selection criteria: 
(1) cores with available interferometric images of the continuum (and some lines) to allow us to select only isolated objects and avoid contamination from nearby sources. 
This condition excluded the HMSCs that are classified as ”warm” in \citet{fontani11}. 
The interferometric continuum images are published in: 
\citet{tan13} for \HMSC, \citet{rivilla20b} for \HMPO,
and \citet{hernandez14} for \UCHII. 
From these images, we could exclude the presence of other significant millimetre continuum sources at an angular distance $\leq 20$\asec\ from the target, larger than the maximum telescope beam radius of our observations ($\sim 16$\asec\ at most).
This allowed us to rule out any significant emission from nearby continuum sources in our spectra.
G034 contains two millimetre cores separated by less than $\sim 15$\asec\ resolved in images of the Atacama Large Millimeter Array \citep[ALMA,][]{tan13}, but they are both starless and hence the source well represents the first stage of the evolutionary sequence shown in Fig.~\ref{fig:sequence}.
(2) Cores that, in previous observations, have the strongest emission in $^{13}$CH$_3$OH (optically thin tracer of surface chemistry, Fontani et al.~\citeyear{fontani15a}) and in DCO$^+$ (3–2) 
(optically thin tracer of gas-phase chemistry, Fontani et al. in prep.). 
Applying these criteria, we selected one target for each evolutionary group. However, among the HMSCs, the selected target would have been G028–C1, which we had to exclude because found to be associated with a young bipolar outflow (Tan et al.~\citeyear{tan16}), and thus with on-going star formation.
Source coordinates, and Local Standard of Rest (LSR) velocities used to centre the 
receiver bands are listed in Table~\ref{tab:sources}. 

\section{Observations and data reduction}
\label{obsandredu}

\subsection{Observations}
\label{obs}

Observations were performed during two observing runs. 
The data of AFGL were taken from the 30th of March to the 2nd of April, 2021 (project 124--20), and those of the other two sources from the 9th to the 19th of April, 2022 (project 129--21).
We covered almost totally the three bands at 3, 2, and 1.3~mm of the Eight MIxer Receiver (EMIR) receiver (hereafter E0, E1, and E2) for the three targets by means of the allowed combinations E0/E2 and E0/E1.
After the observations towards AFGL, the following observations towards G034 and G589 in bands E0, E1, and E2 were optimised to cover some lines not covered in AFGL (for example the $J=8-7$ transition of \CIII).
This results in a marginal difference in the frequency coverage for the three targets, which however is just of $\sim 0.5-2$~GHz depending on the band (Table~\ref{tab:obs}).
Band E3 was observed only towards \HMPO\ through combinations E1/E3 in the first observing run, but it was not observed towards \HMSC\ and \UCHII\ in the following runs because band E2 already showed almost no line towards \HMSC, and it was becoming too dense of blended lines at high frequencies towards \UCHII. 
Each band was covered with individual bands of $\sim 7.78$ GHz (dual polarisation) provided by the fast Fourier transform spectrometer with 200~kHz spectral resolution (FTS200).  

Table~\ref{tab:obs}
shows the observed spectral ranges, as well as some technical details of
the observations in those ranges: the beam full width at half maximum (HPBW), the velocity resolution ($V_{\rm res}$), 
and the telescope beam and forward efficiencies ($B_{\rm eff}$ and $F_{\rm eff}$, respectively) used to convert the spectra from antenna temperature to main beam temperature units.
The observations were made in wobbler-switching mode with a wobbler throw of 220\asec. 
Pointing was checked (almost) every hour on nearby quasars, or bright HII regions. 
Focus was checked on planets at the start of observations, and after sunset and sunrise. 
The data were calibrated with the chopper wheel technique (see Kutner \& Ulich~\citeyear{keu81}), with a calibration uncertainty 
of about $10\%$. 
The goal sensitivity was $10-20$~mK in $T_{\rm a}^*$ units in all spectra.
Because of variable observational (mostly weather) conditions, the final rms is inhomogeneous. 
A representative range of reached $1\sigma$~rms values is given in Table~\ref{tab:obs}.

   \begin{table*}
      \caption[]{Observational parameters.}
         \label{tab:obs}
         \begin{tabular}{lcccccc}
            \hline
            \noalign{\smallskip}
    source    &   observed EMIR band      &  HPBW & $V_{\rm res}$ & $B_{\rm eff}$ & $F_{\rm eff}$ & $1\sigma$ rms\tablefootmark{b} \\
              &   (GHz)     & (\asec , pc)\tablefootmark{c} & (\kms) & &  & (mK) \\
            \noalign{\smallskip}
            \hline
            \noalign{\smallskip}
    G034 (HMSC) & 72.685 -- 117.320 (E0) & $33.7-21\,,\, 0.47-0.30$ & $\sim 0.82 - 0.51$ & 0.82 -- 0.79 & 0.95 -- 0.94 & 5--10 \\
    AFGL (HMPO) & 73 -- 117 (E0) & $33.7-21\,,\,0.29-0.19$ & & & & 6--11 \\
    G589 (UCHII) & 72.685 -- 115.255 (E0) & $33.7-21\,,\,0.21-0.13$ & & & & 6--12 \\
    \noalign{\smallskip}
    \hline
    \noalign{\smallskip}
    G034 (HMSC) &  124.685 -- 182.205 (E1) & $19.7-13.4\,,\,0.28-0.18$ & $\sim 0.48 - 0.33$ & 0.76 -- 0.68 & 0.94 -- 0.93 & 7--14 \\
    AFGL (HMPO) &  125 -- 184 (E1) & $19.7-13.4\,,\,0.17-0.11$ & & & & 11--15 \\
    G589 (UCHII) &  124.685 -- 180.205 (E1) & $19.7-13.4\,,\,0.13-0.08$ & & & & 11--18 \\
    \noalign{\smallskip}
    \hline
    \noalign{\smallskip}
    G034 (HMSC) & 201.685 -- 274.335 (E2) & $12.2-9.0\,,\,0.17-0.13$ & $\sim 0.30-0.22$ & 0.64 -- 0.51 & 0.94 -- 0.88 & 18--32 \\
    AFGL (HMPO) & 202.010 -- 274.010 (E2) & $12.2-9.0\,,\,0.11-0.08$ & & & & 18--32 \\
    G589 (UCHII) & 201.685 -- 274.335 (E2) & $12.2-9.0\,,\,0.08-0.06$ & & & & 18--34 \\
    \noalign{\smallskip}
    \hline
    \noalign{\smallskip}        
    AFGL (HMPO) & 284 -- 347 (E3)\tablefootmark{a} & $8.7-7.1\,,\,0.08-0.06$ & $\sim 0.21 - 0.17$ & 0.48 -- 0.34 & 0.86 -- 0.80 & 45--85 \\
            \noalign{\smallskip}
            \hline
         \end{tabular}
         \tablefoot{\tablefoottext{a}{observed only towards AFGL;}\tablefoottext{b}{in $T_{\rm MB}$ units;}\tablefoottext{c}{half power beam width of the telescope in angular and linear units.}
         }
   \end{table*}

\subsection{Data reduction, line identification, and fitting}
\label{red}

The first steps of data reduction (e.g. average of the scans, baseline removal, flag of 
bad scans and channels) were made with the Continuum and Line Analysis Single-dish Software
({\sc class}) package of the Grenoble Image and Line Data Analysis Software
({\sc gildas}\footnote{https://www.iram.fr/IRAMFR/GILDAS/}) using standard procedures.
Then, the baseline-subtracted spectra in main beam temperature ($T_{\rm MB}$) units were 
imported into the MAdrid Data CUBe Analysis ({\sc MADCUBA}\footnote{{\sc MADCUBA} is a software developed in the Madrid Center of Astrobiology (CAB, CSIC-INTA), which enables to visualise and analyse single spectra and data cubes: https://cab.inta-csic.es/madcuba/.}, Mart\'in et al.~\citeyear{martin19}) software, using the tool {\sc Import Spectra CLASS file}. 

To perform the line identification and fitting we have used the Spectral Line Identification and Modeling ({\sc slim}) tool of {\sc madcuba}, which, among others, contains the Cologne Database for Molecular Spectroscopy (CDMS\footnote{https://cdms.astro.uni-koeln.de/cdms/portal/}; \citealt{endres16}), and the Jet Propulsion Laboratory molecular database \citep[JPL,][]{pickett98} catalogue.
We have generated with {\sc slim} synthetic spectra of the targeted molecules under the assumption of Local Thermodynamic Equilibrium (LTE), and compare with the observed spectra. 
Once the molecule is identified, we have performed the fit applying \textsc{slim-autofit}, which provides the best non-linear least-squares LTE fit to the data using the Levenberg-Marquardt algorithm. 
The parameters used in the LTE model are: molecular column density ($N$), excitation temperature ($T_{\rm ex}$), peak velocity ($V_\text{p}$) and full width at half maximum (FWHM) of the Gaussian line profiles. 
Because we do not know the source size in each tracer, necessary to calculate correctly the beam dilution factor (which could also change for each transition), we have always assumed that the emission fills the telescope beam in each transition (i.e., no beam dilution).
This assumption is also partially justified by the angular extension of the millimeter continuum emission ($\sim 10-20$\asec\ when observed in high-angular resolution images, see reference works in Sect.~\ref{sample}).
However, since the transitions with high $E_{\rm up}$ (e.g. $\geq 100$~K) are supposed to arise from compact regions that could be smaller than the telescope beam, the excitation temperatures derived for molecules including these transitions should be considered as lower limits.
The outcome of \textsc{slim-autofit} gives the value of the parameters, along with their associated uncertainties. 
For some species, \Tex\ had to be fixed either because we did not detect a sufficient amount of lines with different excitation conditions, or because the fit could not converge even with a sufficient number of such lines.
As described in Sect.~\ref{res}, in this paper the species for which we fixed \Tex\ to derive the molecular column densities are \cIIh\ and \cIVh\ towards G034. 
The reasons for the chosen \Tex\ are also explained in Sect.~\ref{res}.

\subsection{Line richness}
\label{richness}

\begin{figure}
\centering
\includegraphics[width=9cm]{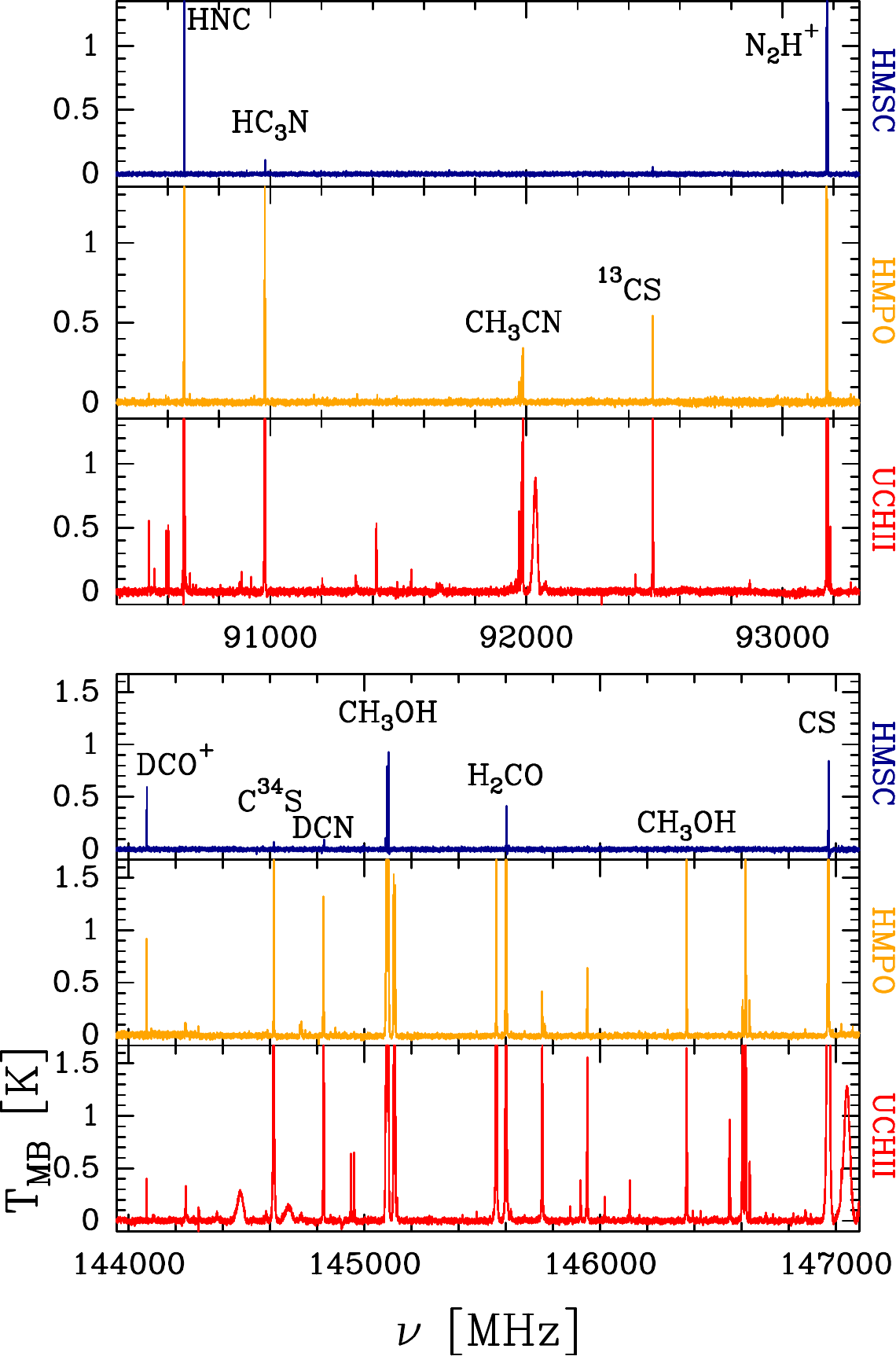}
\caption{\label{fig:spec-comp} Sample spectra of the 3 CHEMICO sources. The two spectral windows are extracted from band E0 and band E1. 
The HMSC \HMSC\ are illustrated as blue histograms, the HMPO \HMPO\ and the UCHII \UCHII\ as orange and red histograms, respectively. 
Relevant molecular lines are indicated.}
\end{figure}

Figure~\ref{fig:spec-comp} shows two spectral windows in the EMIR band E0 and band E1 observed towards the three CHEMICO targets.
Even though this paper does not intend to give a complete picture of the chemical inventory in the three sources, the plot gives the feeling of the different chemical richness.
The number of lines in the spectra increases with evolution, as well as their intensities and broadness.
The plot indicates that the two COMs \METH\ and \cyanide\ show the highest intensity enhancement with evolution, while this is less apparent in the simple species, such as HNC and N$_2$H$^+$.
In the spectra extracted from band E1, we also show the $J=2-1$ line of two deuterated species, DCO$^+$ and DCN.
We notice that while the intensity of the DCN line increases with evolution like that of almost all the other lines, that of DCO$^+$ first increases and then decreases.
This is in agreement with a different dependence of the two molecules with the increasing gas kinetic temperature.
In fact, it is well known that the formation of DCO$^+$ is boosted in cold gas and inhibited with increasing temperature, while the formation of DCN is efficient at higher temperatures \citep[e.g.][]{roueff07,roueff13}.
We also note the detection of molecules containing $^{13}$C and $^{34}$S.
A thorough analysis of the chemical inventory in the three CHEMICO sources will be given in following papers.

\section{Results}
\label{res}

In this work we focus our analysis on \cIIh, \cyclicII, \cyclic, \cIVh, \methyl, \CIII,  \cyanide\ and \CV, that can give us clues on the evolution of carbon-chains and carbon-rich species, and provide estimates of the gas temperature.
The species detected in all targets are: \cIIh, \cyclic, \methyl, and \CIII.
\CV\ and \cyanide\ are both clearly detected towards AFGL and G589 but not towards the HMSC G034.
or \CV, the non-detection can be explained by the fact that the energies of the upper levels of the transitions ($E_{\rm u}$) observable in the EMIR bands start from $\sim 51$~K, and hence they are not easily excited in the earliest cold phase. 
As for the non-detection of \cyanide, even though lines with $E_{\rm u}$ down to $\sim 9$~K could be observed, it is a well-known hot core tracer whose abundance decreases by an order of magnitude from the HMPO phase to the HMSC stage \citep{mininni21}.
We also searched for larger cyanopolyynes (e.g. HC$_7$N), but they are not detected in all sources.

\begin{table*}
    \caption[]{Excitation temperatures derived with {\sc madcuba}. The errors are listed in brackets.
    As reference, in the last column we give \Tk\ derived from the \AMM\ (1,1) and (2,2) inversion transitions analysed by \citet{fontani15a}. 
    Please note that this value refers to a specific layer of the sources.}
         \label{tab:temperatures}
         \begin{tabular}{lccccccccc}
            \hline
            \noalign{\smallskip}
            & \multicolumn{8}{c}{\Tex (K)} & \Tk(K) \\
            \noalign{\smallskip}
            \cline{2-9}
            \noalign{\smallskip}
            & \multicolumn{4}{c}{hydrocarbons} & & \multicolumn{3}{c}{cyanides} & \\
            \cline{2-5} \cline{7-9} \\
Source       & \cIIh & {\it c-}C$_3$H & \cyclic  & \methyl & & \CIII & \cyanide & \CV & \AMM \\
			  \hline
			  \noalign{\smallskip}
G034-1\tablefootmark{a}   & 4\tablefootmark{b} & --    & 4.5(0.4)   & 14.2(1.2) & & 6.8(0.2)  & --          & --              & 15.2(0.4) \\
G034-2                    & 4\tablefootmark{b} & --    & 5.7(0.6)   & --        & & 6.3(0.5)  & --          & --              & \\
AFGL-1\tablefootmark{c}   & 18.0(0.2)          & 14(3) & 17.5(0.4)  & 53.9(0.3) & & 37(0.7)   & --          & 48(2)           & 21.5(0.5) \\
AFGL-2                    & --                 & --    & --         &  --       & & 125(8)    & 117.1(1.4)  & --              & \\
G589-1\tablefootmark{d}   & 18.6(1.1)          & --    & 14.4(1.3)  & 67.2(0.8) & & 58.3(0.3) & 76.8(0.7)   & 120(14)         & 29.0(0.3) \\
G589-2                    & --                 & --    & --         & --        & & 95.6(1.6) & --          & 154(26)         &  \\
            \hline
        \end{tabular}
\tablefoot{\tablefoottext{a}{two velocity features centered at $\sim 41.6$~\kms\ (G034-1) and $\sim 43.3$~\kms\ (G034-2), respectively;}
\tablefoottext{b}{\Tex\ value fixed in {\sc madcuba} for both velocity features, for which the residuals between synthetic and observed spectrum are minimum;}
\tablefoottext{c}{two components with same peak velocity but FWHM of $\sim 2.6$~\kms\ (AFGL-1) and $\sim 6.6$~\kms\ (AFGL-2), respectively;}
\tablefoottext{d}{two components with same peak velocity but FWHM of $\sim 4$~\kms\ (G589-1) and $\sim 11$~\kms\ (G589-2), respectively.}
}
\end{table*}

Table~\ref{tab:temperatures} lists the excitation temperatures (\Tex) of the aforementioned species.
In general, \Tex\ increases with evolution in each species, but in a different way. For example, simple hydrocarbons show an increase by a factor $\sim 3-4$ from the HMSC and the HMPO/UCHII stages, while cyanopolyynes are associated with \Tex\ enhancements up to a factor $\sim 10$.
A detailed comparison among the various excitation temperatures will be made in Sect.~\ref{discu-tex}.
The lines detected with the highest signal to noise ratio are those of \CIII.
Table~\ref{tab:lines} lists the rotational transitions of HC$_3$N that are observable in the EMIR bands. 
We used this species as a reference for the most accurate temperature estimate owing to the huge number of transitions detected, covering a range of $E_{\rm u}$ of $\sim 20 - 300$~K.
Some lines could be optically thick, but in the fit of the spectra {\sc madcuba} takes opacity of the transitions into consideration.
Towards G034, we have detected up to the $J=12-11$ transition in G034; towards AFGL, up to the $J=36-35$ transition; towards G589, up to the $J=30-29$. 
But, as previously said, we did not observe the band E3 lines towards G589.
We show the total molecular column densities, obtained from the best (Gaussian) fit to the lines as described in Sect.~\ref{red}, in Table~\ref{tab:columns}.
The best fit peak velocities ($V_{\rm p}$) and full widths at half maximum (FWHM) are listed in Tables~\ref{tab:fit-v} and \ref{tab:fit-fwhm}, respectively, of Appendix~\ref{app:fits}.

   \begin{table}
      \caption[]{Rotational lines of HC$_3$N observable in the 4 EMIR bands. In the last three columns, we list if the transition has been detected (Y) or undetected (N) in each core.}
         \label{tab:lines}
         \setlength{\tabcolsep}{1.2pt}
         \begin{tabular}{lccccccc}
            \hline
            \noalign{\smallskip}
            Rest freq.\tablefootmark{a} &       $J_{\rm u}-J_{\rm l}$ & $S\mu_{ul}^2$\tablefootmark{a} & $E_{\rm u}$\tablefootmark{a} & $A_{\rm ul}$\tablefootmark{a} & HMSC & HMPO & UCHII \\
               (GHz)     &  & (Dy$^2$) & (K) & (s$^{-1}$) & & & \\
            \noalign{\smallskip}
            \hline
            \noalign{\smallskip}
     72.7823  &  8--7    &  111 & 15.7  &   2.9e-05 & Y & --\tablefootmark{b} & Y \\
     81.8815  &   9--8    & 125 & 19.6  &	  4.2e-05 & Y & Y & Y \\	
     90.9790  &  10--9   & 139 & 24.0  &	  5.8e-05 & Y & Y & Y \\	
    100.0764 &  11--10   & 153 & 28.8  &	  7.7e-05 & Y & Y & Y \\		
    109.1736 &  12--11 & 166 & 34.1  &	  0.00010 & Y & Y & Y \\
    127.3677 &  14--13 & 194 & 45.8  &	   0.00016 & N & Y & Y  \\	
    136.4644 &  15--14 & 208 & 52.4  &	  0.00020 & N & Y & Y \\	
    145.5610 &  16--15 & 222 & 59.4  &	  0.00024 & N & Y & Y \\	
    154.6573 &  17--16 & 236 & 66.8  &	 0.00029 & N & Y & Y \\		
    163.7534 &  18--17 & 250 & 74.7  &	 0.00034 & N & Y & Y \\		
    172.8493 &  19--18 & 264 & 83.0  &	  0.00041 & N & Y & Y \\	
    181.9449 &  20--19 & 277 & 91.7  &	  0.00047 & N & Y & -- \\	
    209.2302 &  23--22 & 319 & 120.5 &    0.00072 & N & Y & Y \\
    218.3248 &  24--23 & 333 & 131.0 &    0.00082 & N & Y & Y \\
    227.4189 &  25--24 & 347 & 141.9 &    0.00093 & N & Y & Y \\
    236.5128 &  26--25 & 361 & 153.2 &     0.00105 & N & Y & Y \\
    245.6063 &  27--26 & 374 & 165.0 &     0.00117 & N & Y & Y \\
    254.6995 &  28--27 & 388 & 177.3 &      0.00131 & N & Y & Y \\
    263.7923 &  29--28 & 402 & 189.9 &     0.00146 & N & Y & Y \\
    272.8847 &  30--29 & 416 & 203.0 &     0.00161 & N & Y & Y \\
    291.0684 &  32--31 & 444 & 230.5 &     0.00196 & -- & Y & -- \\
    300.1597 &  33--32 & 458 & 244.9 &     0.00215 & -- & Y & -- \\
    309.2504 &  34--33 & 472 & 259.8 &     0.00235 & -- & Y & -- \\
    318.3408 &  35--34 & 485 & 275.0 &     0.00257 & -- & Y & -- \\
    327.4307 &  36--35 & 499 & 290.8 &     0.00279 & -- & Y & -- \\
    336.5201 &  37--36 & 513 & 306.9 &     0.00304 & -- & Y & -- \\
    345.6090 &  38--37 & 527 & 323.5 &     0.00329 & -- & Y & -- \\
	           \noalign{\smallskip}
            \hline
         \end{tabular}
    \tablefoot{\tablefoottext{a}{from the CDMS catalogue;}
    \tablefoottext{b}{not observed.}
    }
\end{table}

\subsection{Line parameters and \Tex\ in the HMSC G034}
\label{g034}

\begin{figure}
    \centering
    \includegraphics[width=1\linewidth]{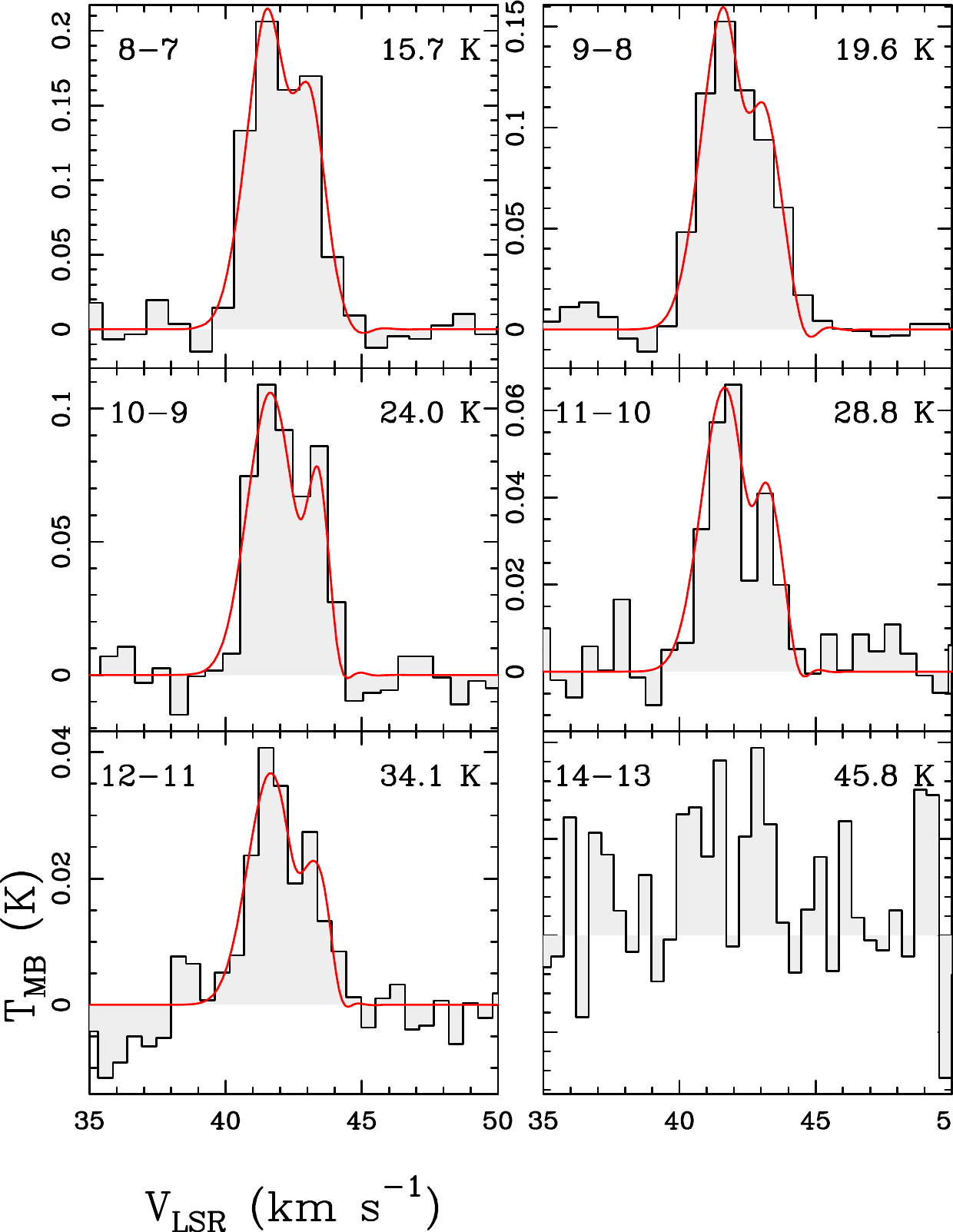}
    \caption{Spectra of the \CIII\ lines detected towards the HMSC, G034. The red curve corresponds to the best fit obtained with {\sc madcuba} using two velocity components.
    Rotational quantum numbers and $E_{\rm u}$ of each transition are indicated in the top left and right corners, respectively, in each frame.}
    \label{fig:HC3N-G034}
\end{figure}

In G034, as it can be seen from the spectra in Fig.~\ref{fig:HC3N-G034}, the \CIII\ lines are well-fitted by two velocity features having similar FWHM ($\sim 1-1.5$~\kms) but different $V_{\rm LSR}$.
The presence of two velocity components in G034 was already noticed in the HCN isotopologues \citep{colzi18a}, and in the \AMM\ (1,1) and (2,2) inversion transitions \citep{fontani15a}, and it is due to the two starless cores identified in the ALMA images \citep{tan13} not resolved in the IRAM 30m beam (as discussed in Sect.~\ref{sample}).
Because both the FWHMs and the separation in velocity of the two components is similar in all \CIII\ lines, and it is revealed also in \cIIh, \cyclic, and \cIVh, it is natural to explain these features as two cores embedded within the IRAM 30m beam.

The \Tex\ measured from \CIII\ and \cyclic\ are $6.8\pm 0.2$~K and $4.5\pm 0.4$~K, respectively, for the velocity component centred at $\sim 41.3$~\kms, and $6.3\pm 0.5$ and $5.7\pm 0.6$ K, respectively, for the velocity component centred at $\sim 43.3$~\kms.
These values indicate emission of these molecules from cold and quiescent gas.
For \methyl, the two velocity components are not resolved, which
means that the FWHM of each component is larger than (or comparable to) the separation in velocity of the two components (2~\kms).
Therefore, the \methyl\ emission is associated with gas warmer and more turbulent than that responsible for the \CIII\ and \cyclic\ emission.
In fact, the \Tex\ measured in \methyl\ is $14.2\pm 1.2$~K.
This value is also in good agreement with \Tk\ estimated from \AMM\ (Table~\ref{tab:temperatures}), indicating that \methyl\ can be in LTE conditions, while the low \Tex\ derived from \CIII\ and \cyclic\ could indicate that these species arise predominantly from a diffuse low-density envelope, where their lines are likely sub-thermally excited.
In fact, because both molecules have large dipole moments \citep[e.g.][]{shirley15}, their critical densities are relatively high ($\simeq 10^5-10^6$~\cmc).
We also estimated \Tex\ for \cIIh, even though in this case the fit to the two velocity features could not converge if we leave both \Tex\ and the total column density as free parameters.
Therefore, to derive \Ntot, we have created synthetic spectra fixing \Tex\ to an integer value in between 3 and 15 K, namely  between the cosmic microwave background temperature and the kinetic temperature derived from ammonia (Table~\ref{tab:temperatures}, and found that 4~K is the value that gives us the lowest residuals between the synthetic spectrum and the data.

\subsection{Line parameters and \Tex\ in the HMPO AFGL}
\label{afgl}

\begin{figure}
   \centering
   \includegraphics[width=8.2cm]{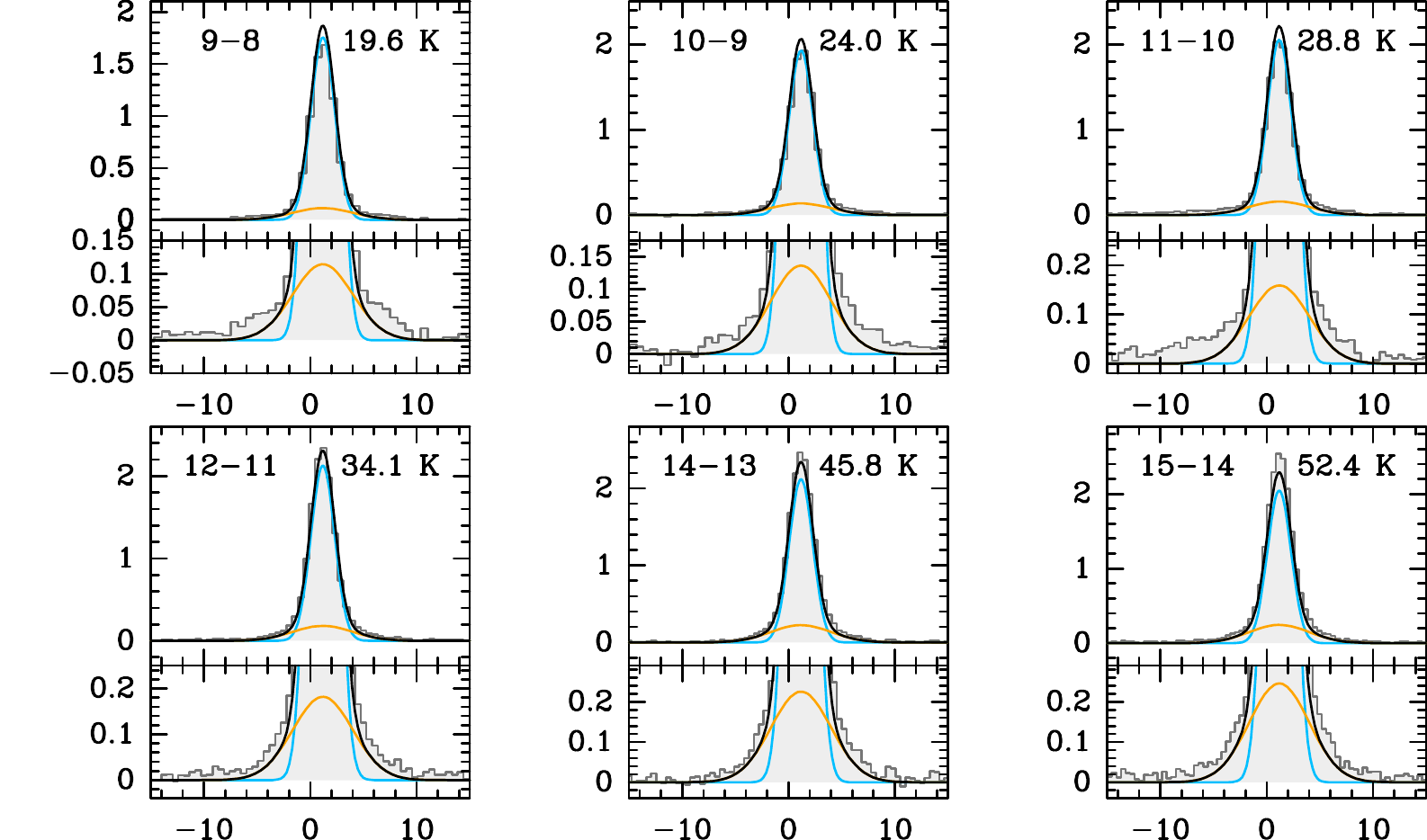}
   \includegraphics[width=8.2cm]{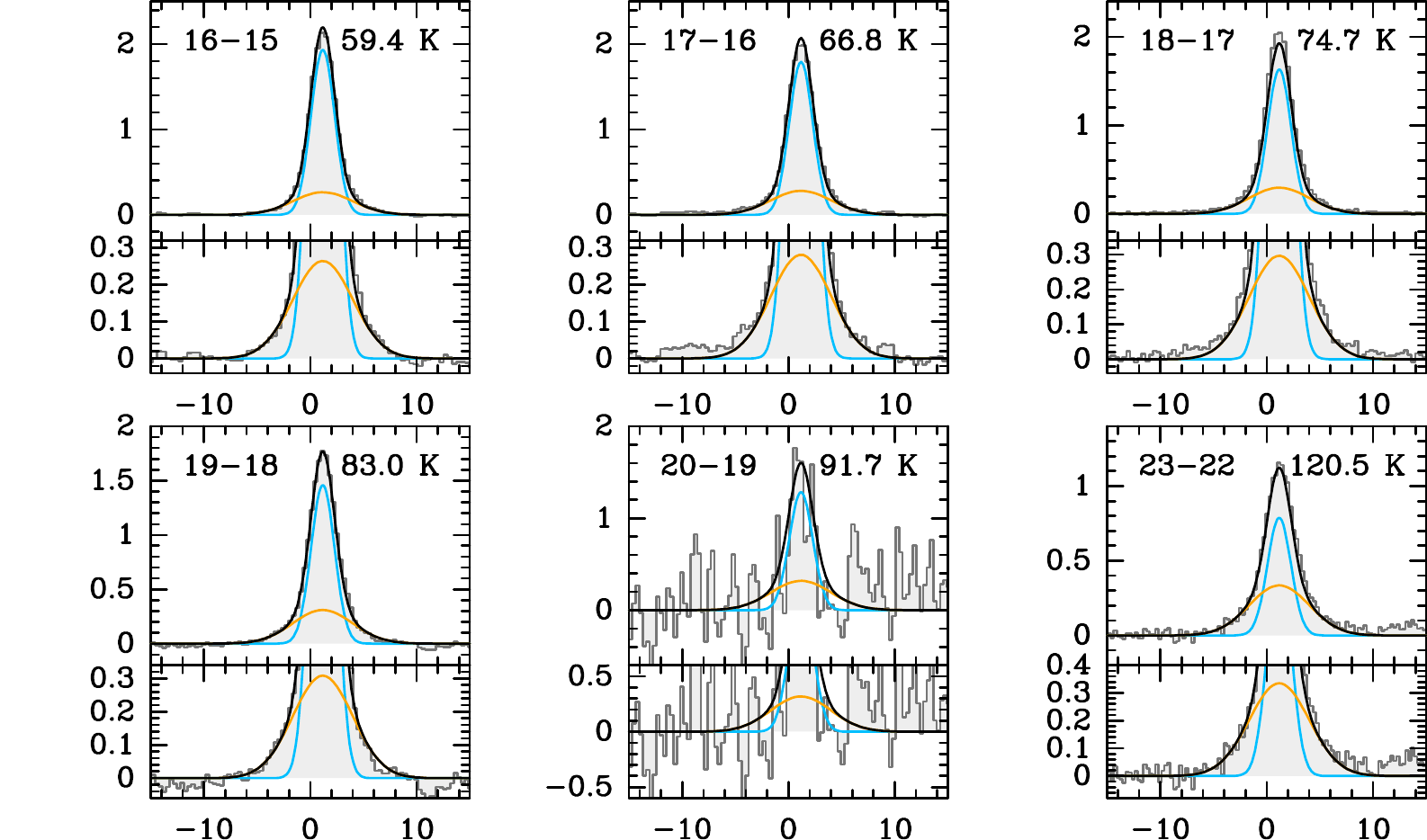}
   \includegraphics[width=8.2cm]{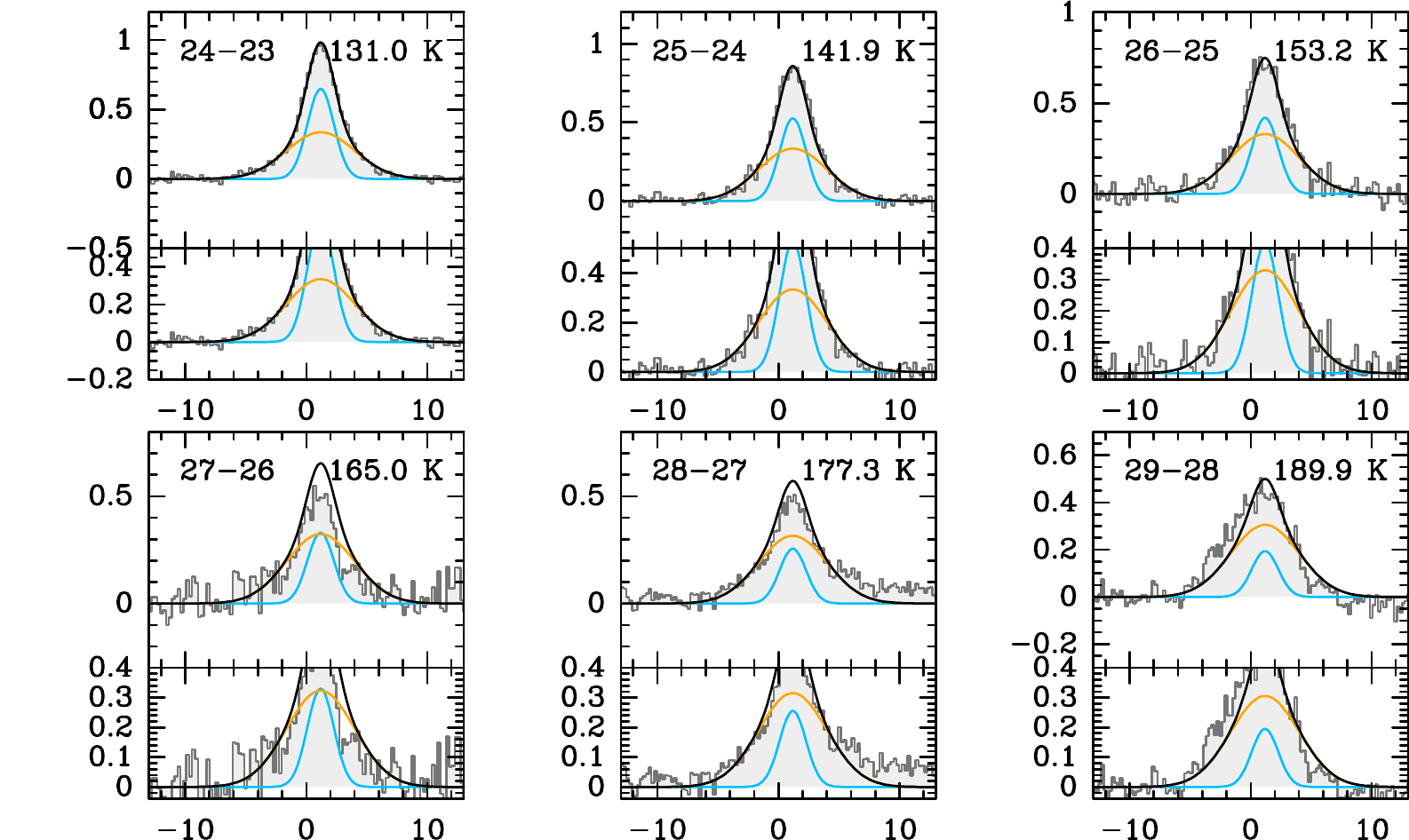}
   \includegraphics[width=8.2cm]{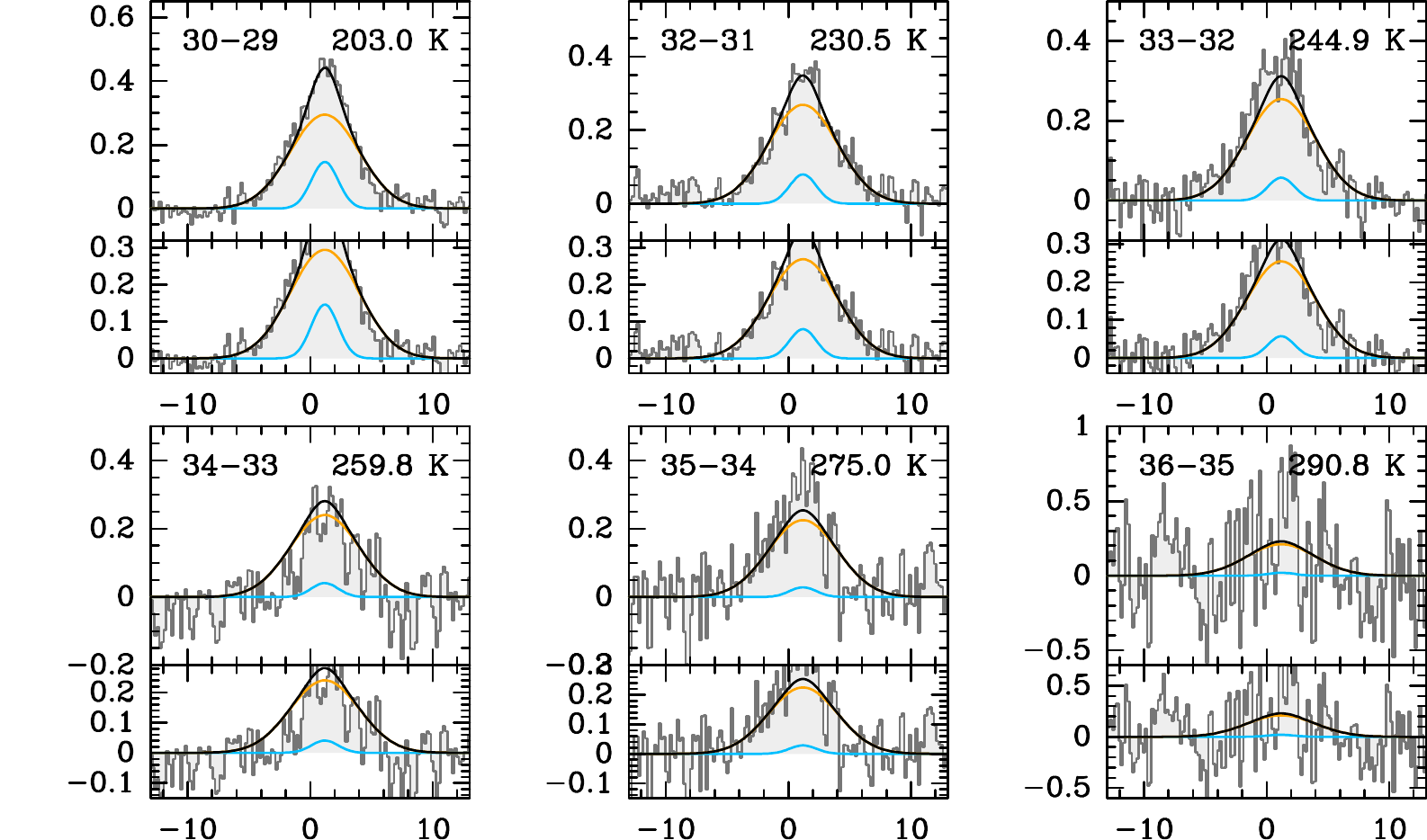}
   \includegraphics[width=8.2cm]{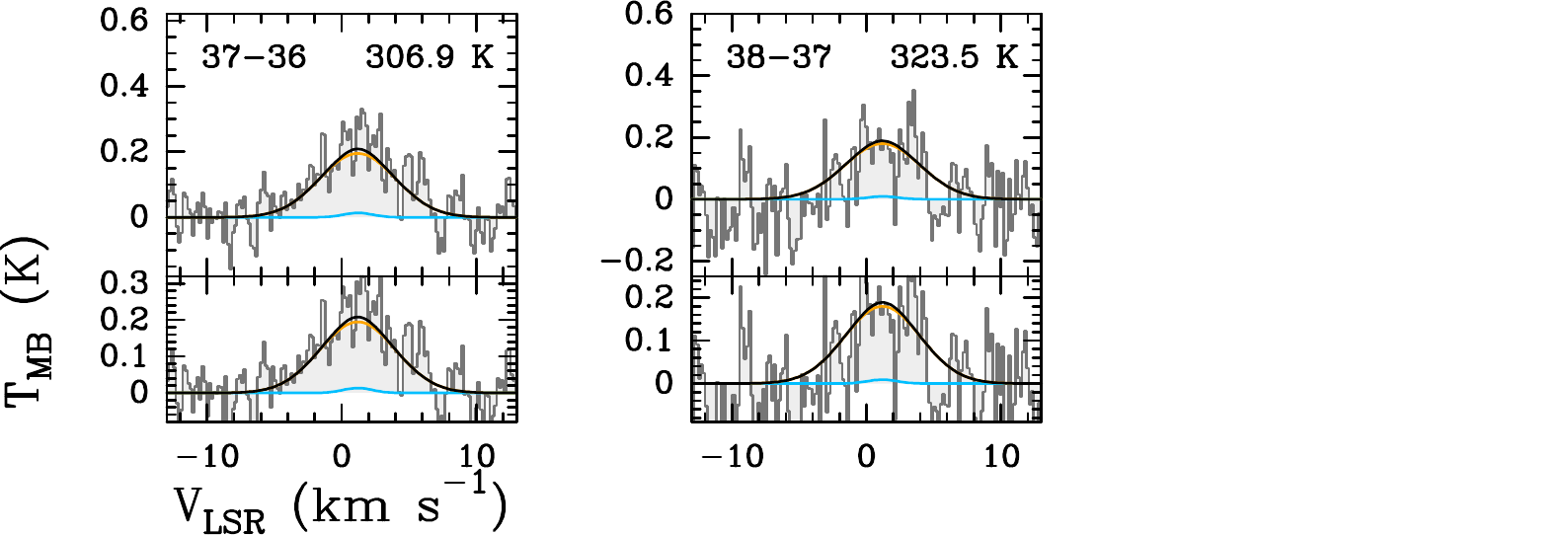}
      \caption{Spectra of the HC$_3$N lines detected towards the HMPO, \HMPO. 
      In each frame, the best Gaussian fit 
      to the narrow and broad velocity components are illustrated by the blue and orange curves, respectively, while the black one is their sum. 
      The frames at the bottom of each spectrum is a zoom on the Y-axis to highlight the high-velocity wings.
      The rotational quantum numbers and $E_{\rm u}$ of each transition are given in the top-left and top right corner, respectively.
              }
         \label{fig:spectra-HMPO}
   \end{figure}
   
\begin{figure}
\centering
   \includegraphics[width=9cm]{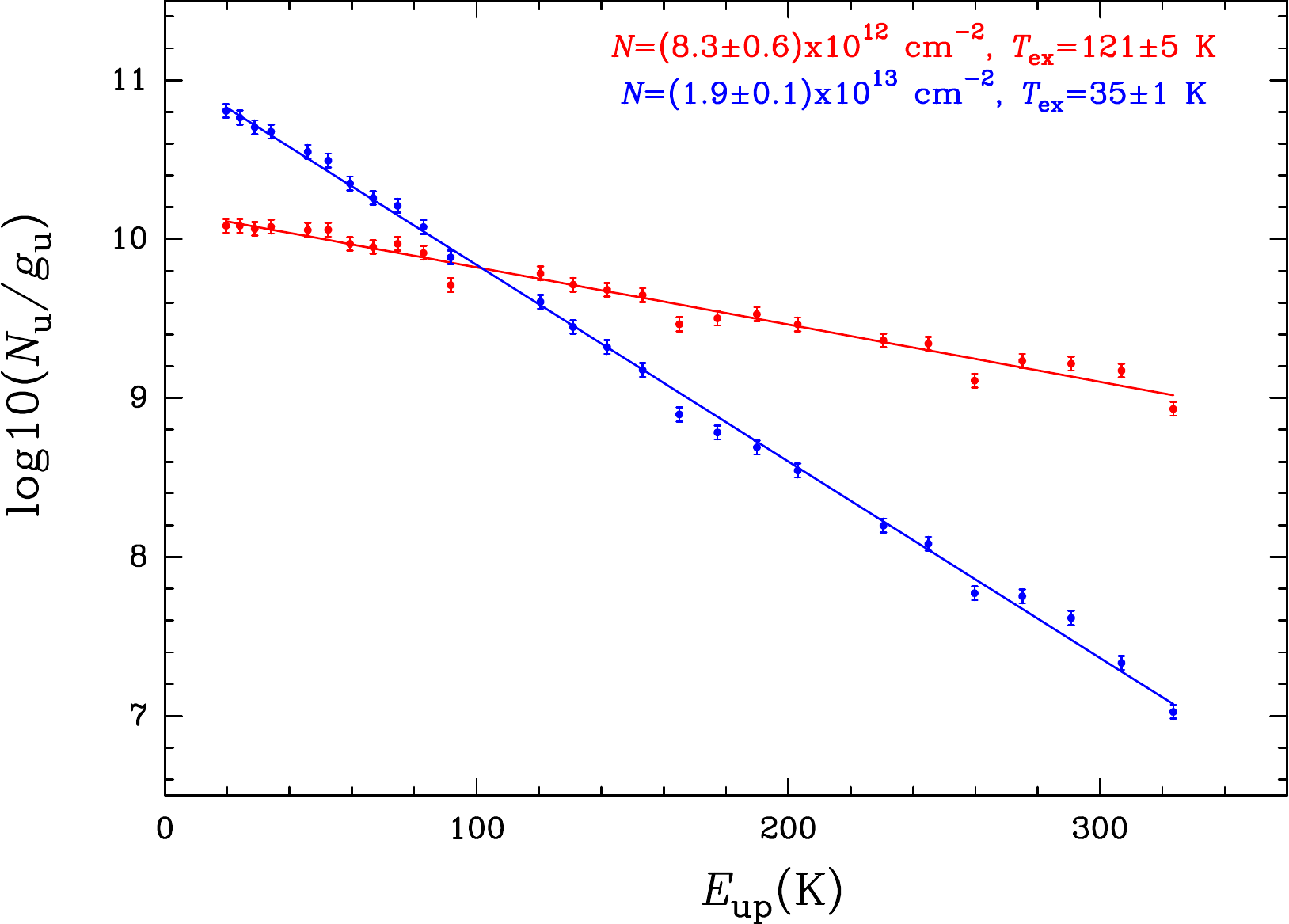}
      \caption{Rotational diagram obtained with {\sc madcuba} towards AFGL (HMSO) from the detected HC$_3$N lines. 
      The The blue and red lines indicate the narrow (cold) and broad (hot) velocity features, respectively.
      The errorbars on the y-axis contains the uncertainties calculated by the fit procedure and the calibration error of $10\%$ on the line integrated intensity.}
         \label{fig:rotation}
   \end{figure}

Figure~\ref{fig:spectra-HMPO} shows the \CIII\ lines detected toward \HMPO.
Each line is well fitted by two Gaussian features centred at similar velocity: a "narrow" one having a FWHM of about 3\kms\ 
that dominates the profile in the lines with $E_{\rm u}$ up to $\sim 120$~K, 
and a "broad" one with FWHM $\sim 7$~\kms\ that becomes more and more prominent in transitions at higher energies.
To fit the lines, we used the CDMS spectral parameters, which take the hyperfine structure into consideration.
However, the velocity separation among the relevant hyperfine components (i.e. those with Einstein coefficient of spontaneous emission, $A_{\rm ul}$, $\geq 10^{-5}$~s$^{-1}$) is at most $\sim 1$~\kms, namely smaller than the FWHM of the lines.
Figure~\ref{fig:spectra-HMPO} shows the spectra with the best fit to the two features, and their sum.
The narrow component is detected up to the $J=32-31$ line, and it has a FWHM comparable to that of all hydrocarbons and \CV\ (Table~\ref{tab:fit-fwhm}), while
the broad component is clearly detected in all transitions of \CIII\ (Table~\ref{tab:fit-fwhm}).
The lack of such a broad component in all the other molecular species can be due either to an insufficient sensitivity to such fainter emission or to the fact that the other species are not appropriate to trace this gas.
As it can be seen in Fig.~\ref{fig:rotation}, the data of the two velocity components are both in excellent agreement with a linear relation, as expected in LTE, but with different \Tex:
the narrow \CIII\ component arises from gas colder (\Tex$\sim 37$~K, Table~\ref{tab:temperatures}) than that associated with the broad one (\Tex$\sim 125$~K).
In particular, \Tex\ of the broad \CIII\ component is in agreement within the errors with that derived from \cyanide\ ($\sim 117$~K, Table~\ref{tab:temperatures}).
                                              
The \cyanide\ and broad \CIII\ emissions have a very similar FWHM ($\sim 6-7$~\kms, Fig.~\ref{fig:Dv}), and they both likely arise from 
the central region of AFGL, which harbours several millimetre cores, some of which are characterised by hot core-like temperatures \citep[e.g.][]{zhang07,rivilla20b}.
\citet{rivilla20b} also detected an extended outflow that can contribute to the emission in this broad component.
Inspecting the other \Tex\ estimates, the three simple hydrocarbons \cyclic, \cyclicII\ and \cIIh\ have \Tex\ in between $\sim 14-18$~K, while the narrow component of \CIII, \CV, and \methyl\ have \Tex\ in between $\sim 37-54$~K.
In general, the lines that can be observed in the IRAM bands are associated with an energy range that is specific of each species: 
smaller molecules are observable in transitions that are associated with energy ranges lower than those observable in the larger ones.
For example, the $E_{\rm up}$ range of the transitions that can be observed in \cIIh\ is $\sim 4-25$~K; in \cIVh\ is $\sim 17-185$~K ($\sim 17-304$~K in AFGL); in \CIII\ is $\sim 16-203$~K ($\sim 20-323$~K in AFGL); in \methyl\ is $\sim 12-696$~K ($\sim 12-2762$~K in AFGL); in \cyanide\ is $\sim 9-670$~K ($\sim 9-2204$~K in AFGL); in \CV\ is $\sim 52-684$~K ($\sim 52-1088$~K in AFGL).
We give the range of the observable and detected transitions in each target in Table~\ref{tab:energies}.
However, the lines detected in the CHEMICO dataset span a broad range of overlapping energies (Table~\ref{tab:energies}).
Therefore, the different \Tex\ derived for hydrocarbons and cyanides depends only partially on the different energy range probed by the transitions observed.

\begin{table*}[]
    \centering
    \caption{Range of observed and detected $E_{\rm u}$ for each species in each source. 
    We do not include \CIII, whose observed and detected transitions are already shown in Table~\ref{tab:lines}.}
    \begin{tabular}{lllllllll}
    \hline
    \noalign{\smallskip}
     &  \multicolumn{8}{c}{$E_{\rm u,min}(K)-E_{\rm u,max}(K)$\tablefootmark{a}} \\
    \noalign{\smallskip}
Species   & \multicolumn{2}{c}{G034(HMSC)} & & \multicolumn{2}{c}{AFGL(HMPO)} & & \multicolumn{2}{c}{G589(UCHII)} \\
   \cline{2-3} \cline{5-6} \cline{8-9} \\
          & observed & detected & & observed & detected & & observed & detected \\
    \hline
    \noalign{\smallskip}
\cIIh     & 4.2 -- 25.2 & 4.2 -- 12.6  & & 4.2 -- 25.2  & 4.2 -- 25.2 & & 4.2 -- 25.2 & 4.2 -- 25.2 \\
\cyclicII & 4.4 -- 345  & --           & & 4.4 -- 345   & 4.4 -- 29   & & 4.4 -- 345  & --          \\
\cyclic\tablefootmark{b}   & 6.4 -- 298  & 6.4 -- 19.3  & & 6.4 -- 298   & 6.4 -- 47   & & 6.4 -- 298  & 6.4 -- 77.2 \\
\cIVh     & 16.5 -- 185 & 16.5 -- 25.1 & & 16.5 -- 304  & 16.5 -- 30  & & 16.5 -- 185 & --          \\
\methyl   & 12.3 -- 696 & 12.3 -- 36.9 & & 12.3 -- 2762 & 12.3 -- 237 & & 12.3 -- 696 & 12.3 -- 372 \\
\cyanide  & 8.8 -- 670  & --           & & 8.8 -- 2204  & 8.8 -- 329  & & 8.8 -- 670  & 8.8 -- 549  \\
\CV       & 52 -- 684   & --           & & 52 -- 1088   & 52 -- 150   & & 52 -- 684   & 52 -- 242   \\
\hline
    \end{tabular}
    \tablefoot{\tablefoottext{a}{Minimum ($E_{\rm u, min}$) and maximum ($E_{\rm u, max}$) energy of the upper level of the transitions, respectively;}
    \tablefoottext{b}{we list only transitions with $E_{\rm u}\leq 300$~K, as the detected transitions have much lower energies in all sources;}
}
\label{tab:energies}
\end{table*}

\begin{figure*}
   \centering
   \includegraphics[width=17cm]{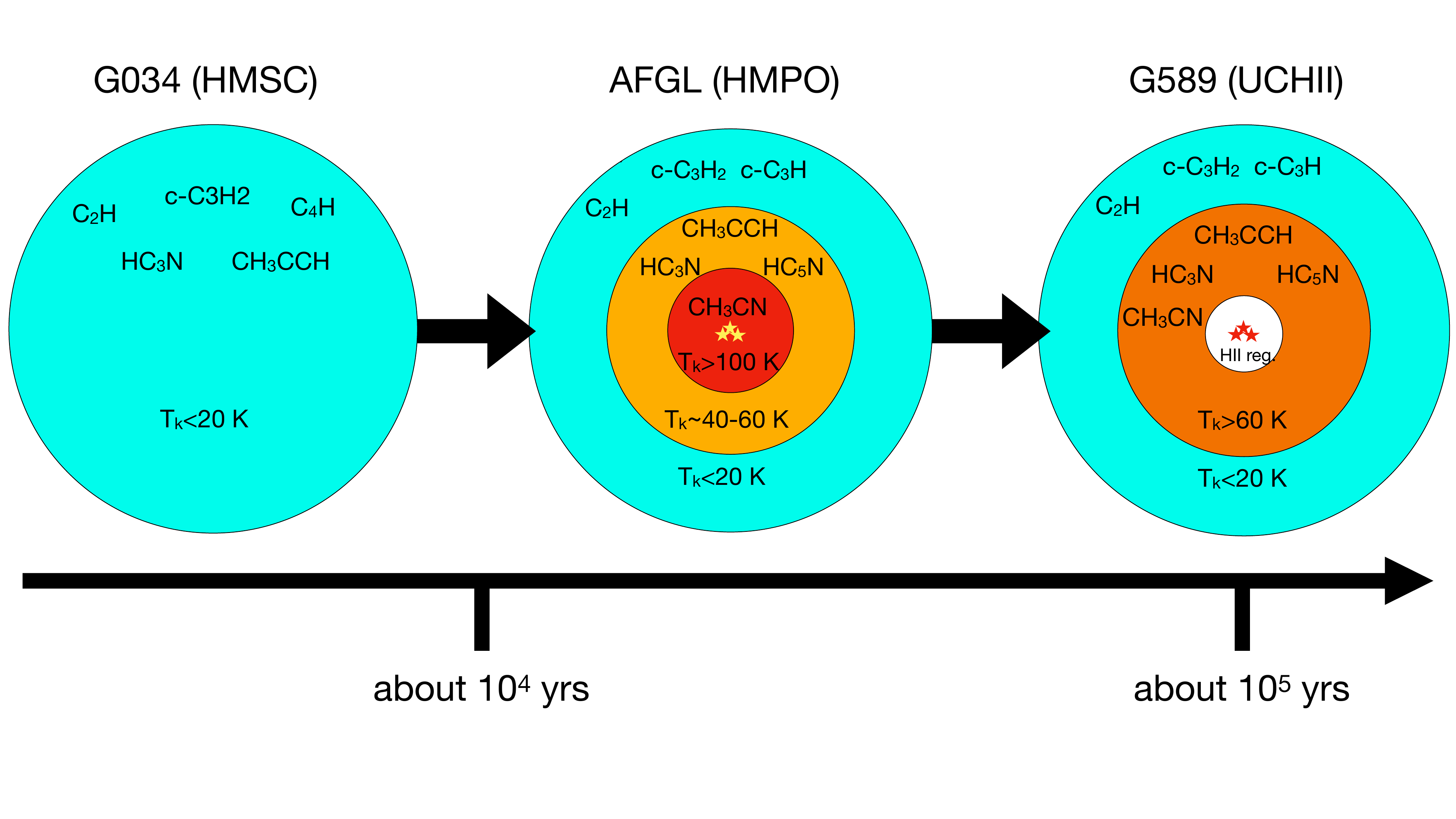}
      \caption{Sketch of the temperature structure in G034, AFGL and G589.
      The molecules studied in this work that preferentially trace each layer are indicated.}
         \label{fig:AFGL}
   \end{figure*}

The different \Tex\ values in AFGL suggests a temperature structure which could be approximated with layers at different temperatures: a cold envelope traced by simple carbon chains with \Tk$\leq 20$~K; a hot core region with \Tk$\geq 100$~K; an intermediate warm region in between the cold envelope and the hot core with \Tk$\sim 40-50$~K.
The diversification of the temperature suggested by our results is illustrated in Figure~\ref{fig:AFGL}, where we highlight the difference from the earlier HMSC phase, characterised by a uniform temperature.
Higher angular resolution observations of the carbon-bearing species studied are needed to unveil the details of this structure, and quantify the radial extension of each layer.
   
\subsection{Line parameters and \Tex\ in the UCHII G589}
\label{g589}

\begin{figure}
    \centering
    \includegraphics[width=1\linewidth]{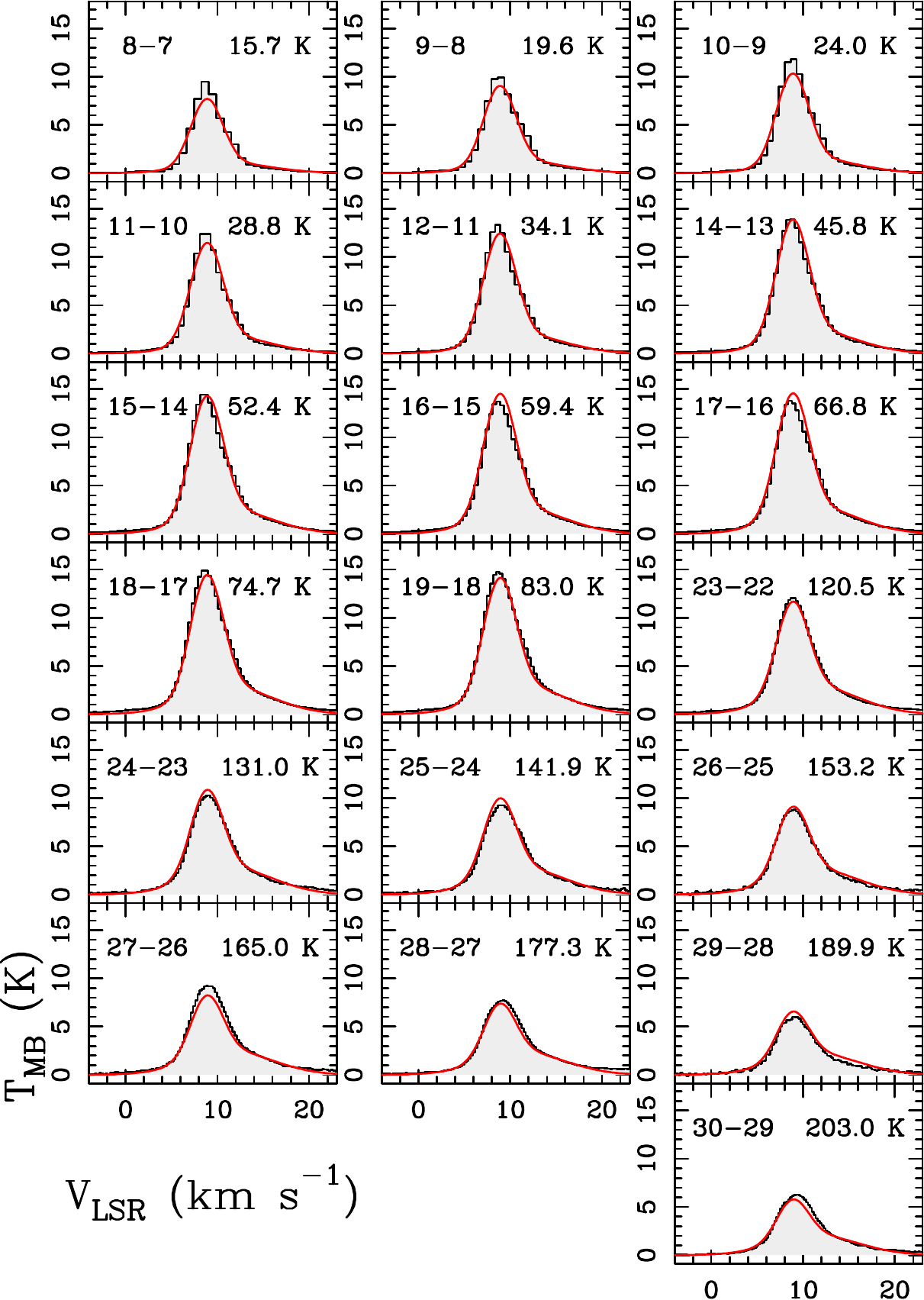}
    \caption{Spectra of the \CIII\ lines detected towards the UCHII, G589. The red curve corresponds to the best fit obtained with {\sc madcuba} using two velocity components.
    The rotational quantum numbers and the energy of the upper level of each transition are indicated in the top left and right corners, respectively.}
    \label{fig:HC3N-G589}
\end{figure}

\begin{figure}
\centering
   \includegraphics[width=9cm]{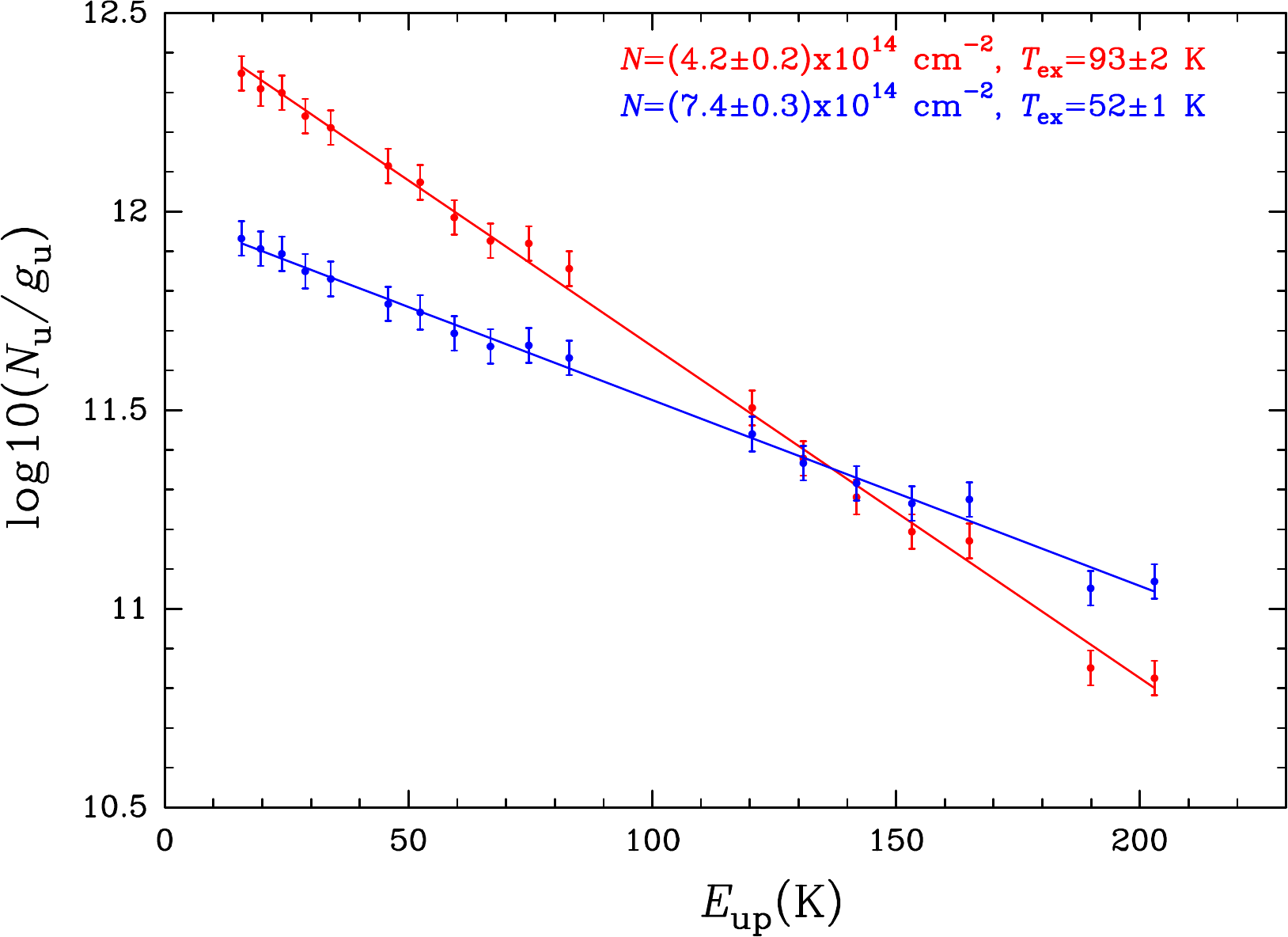}
      \caption{Rotational diagram obtained with madcuba towards G589 (UCHII) from the detected \CIII\ lines. 
      The blue and red lines indicate the
narrow and broad velocity features, respectively.
The errorbars on the y-axis are calculated as explained in the caption of Fig.~\ref{fig:rotation}}.
         \label{fig:rotation2}
   \end{figure}

The lines of \CIII\ detected towards \UCHII\ are shown in Fig.~\ref{fig:HC3N-G589}.
This target shows the most intense transitions.
As for \HMPO, the \CIII\ lines are better fitted when two velocity features, a narrower one ($\sim 4$~\kms) and a broader one ($\sim 11$~\kms), are considered.
Fig.~\ref{fig:rotation2} shows the rotational diagram obtained by fitting separately these two velocity components.
The narrower component has a FWHM comparable to that of all hydrocarbons (Table~\ref{tab:fit-fwhm}), as in \HMPO, but also to that of \cyanide.
This difference indicates that the gas turbulence is progressively increasing.
The broad \CIII\ component is likely associated with the powerful outflow driven by the source(s) embedded in the centre of the UCHII region \citep{zapata20}.

\subsection{Molecular column densities and fractional abundances}
\label{abundances}

As described in Sect.~\ref{red}, we analysed the spectra through a LTE approach that provided both \Tex\ and the total molecular column densities, \Ntot. 
These are listed in Table~\ref{tab:columns}.
In each species detected, \Ntot\ increases with evolution.
The species associated with the highest \Ntot\ enhancement (by two or more orders of magnitude) from the HMSC to the later stages are \cyanide\ and \methyl.
From the $N$(H$_2$) column densities listed in Table~\ref{tab:sources}, we have also derived the molecular fractional abundances, $X$, by dividing \Ntot\ by $N$(H$_2$).
All $X$ are listed in Table~\ref{tab:abb}.
Because $N$(H$_2$) is an average value in an angular region of $28$\asec, and \Ntot\ is computed assuming that the emission fills the beam of the CHEMICO observations, $X$ is also an average value over $28$\asec.
In case of two velocity components, we have summed the column densities and obtained $X$ by dividing the sum for $N$(H$_2$).
The most abundant species are \cIIh\ and \methyl, both having $X$ of the order of $10^{-9}$.
All the other species have $X$ in the range $10^{-10}-10^{-11}$.
A detailed comparison between the fractional abundances and a relation with the evolutionary stage of the sources will be made in Sect.~\ref{discu-abb}.

\begin{table*}
    \caption[]{Total column densities derived with {\sc madcuba}. The errors from the {\sc madcuba-autofit} fitting are listed in brackets.}
         \label{tab:columns}
         \begin{tabular}{lccccccccc}
            \hline
            \noalign{\smallskip}
            Source & \multicolumn{9}{c}{\Ntot (\cmq)} \\
            \noalign{\smallskip}
                 & \multicolumn{5}{c}{hydrocarbons} & & \multicolumn{3}{c}{cyanides} \\
            \noalign{\smallskip}
            \cline{2-6} \cline{8-10} \\
          & \cIIh  & {\it c-}C$_3$H & \cyclic & \cIVh & \methyl & & \CIII\ & \cyanide & \CV\ \\
            \noalign{\smallskip}
            &  $\times 10^{12}$ & $\times 10^{12}$ & $\times 10^{12}$ & $\times 10^{12}$ & $\times 10^{12}$ & & $\times 10^{12}$ & $\times 10^{12}$ & $\times 10^{12}$ 
            \\
            \hline
            \noalign{\smallskip}
            G034-1\tablefootmark{a} & 43(4)     & $\leq 0.8$ & 4.8(0.5) & 9.8(0.9)\tablefootmark{d} & 8.7(0.6)  & & 4.0(0.3) & $\leq 0.7$ & $\leq 0.4$ \\
            G034-2    & 25(3)     & --         & 2.9(0.3) & 5.6(0.5)\tablefootmark{d} & --        & & 1.4(0.3) & --         & --         \\
            AFGL-1\tablefootmark{b} & 501(3)    & 1.4(0.3)   & 6.0(0.2) & $\leq 0.6$                & 468(3)    & & 24.0(0.5)& --  & 1.8(0.1)   \\
            AFGL-2     & --        & --         & --       & --                        & --        & & 8.9(0.5) & 26.9(0.2)         & --         \\
            G589-1\tablefootmark{c} & 1860(100) & $\leq 4$   & 30(3)    & $\leq 2.5$                & 5100(60)  & & 229(2)   & 147(1)     & 10.5(1.2)  \\
            G589-2                  & --        & --         & --       & --                        & --        & & 126(2)   & --         & 6.0(1.1)   \\
            \hline
        \end{tabular}
\tablefoot{\tablefoottext{a}{two velocity features centred at $\sim 41.6$~\kms\ (G034-1) and $\sim 43.3$~\kms\ (G034-2), respectively;}
\tablefoottext{b}{two components with same peak velocity but FWHM of $\sim 2.6$~\kms\ (AFGL-1) and $\sim 6.6$~\kms\ (AFGL-2), respectively;}
\tablefoottext{c}{two components with same peak velocity but FWHM of $\sim 4$~\kms\ (G589-1) and $\sim 11$~\kms\ (G589-2), respectively;}
\tablefoottext{d}{Computed assuming \Tex\ from \cyclic;}}
\end{table*}

\begin{table*}
    \caption[]{Molecular fractional abundances with respect to H$_2$. The errors are listed in brackets and include a $10\%$ calibration error on the molecular column densities.}
         \label{tab:abb}
         \begin{tabular}{lccccccccc}
            \hline
                       \noalign{\smallskip}
            Source & \multicolumn{9}{c}{Molecular abundances} \\
            \noalign{\smallskip}
             & \multicolumn{5}{c}{hydrocarbons} & & \multicolumn{3}{c}{cyanides} \\
            \noalign{\smallskip}
            \cline{2-6} \cline{8-10} \\
            & \cIIh & {\it c-}C$_3$H & \cyclic & \cIVh & \methyl & & \CIII & \cyanide & \CV \\
			 \noalign{\smallskip}
			 & $\times 10^{-9}$ & $\times 10^{-11}$ & $\times 10^{-10}$ & $\times 10^{-10}$ & $\times 10^{-9}$ & & $\times 10^{-10}$ & $\times 10^{-10}$ & $\times 10^{-11}$  \\
			 \hline
			 \noalign{\smallskip}
          G034 (HMSC)\tablefootmark{a} & 1.6(0.5)  & $\leq 2$   & 1.9(0.5) & 3.8(0.9)\tablefootmark{b} & 0.21(0.09) & & 1.3(0.4) & $\leq 0.1$  & $\leq 1$\\   
        AFGL (HMPO)\tablefootmark{a} & 5(2)      & 1.4(0.8)   & 0.6(0.2) & $\leq 0.06$               & 4.7(1.6)   & & 3.3(0.9) & 2.7(0.9)   & 1.8(0.6)\\
        G589 (UCHII)\tablefootmark{a} & 3.4(1.4)  & $\leq 0.7$ & 0.5(0.2) & $\leq 0.05$               & 9(3)       & & 6.5(1.9) & 2.7(1.0)    & 3.0(0.9)\\
            \hline
        \end{tabular}
\tablefoot{\tablefoottext{a}{computed summing the column densities of the two velocity features, when present;}
\tablefoottext{b}{Computed assuming \Tex\ from \cyclic;}}
\end{table*}

\section{Discussion}
\label{discu}

\subsection{Evolution of the excitation temperatures}
\label{discu-tex}

\begin{figure}
    \centering
    \includegraphics[width=1\linewidth]{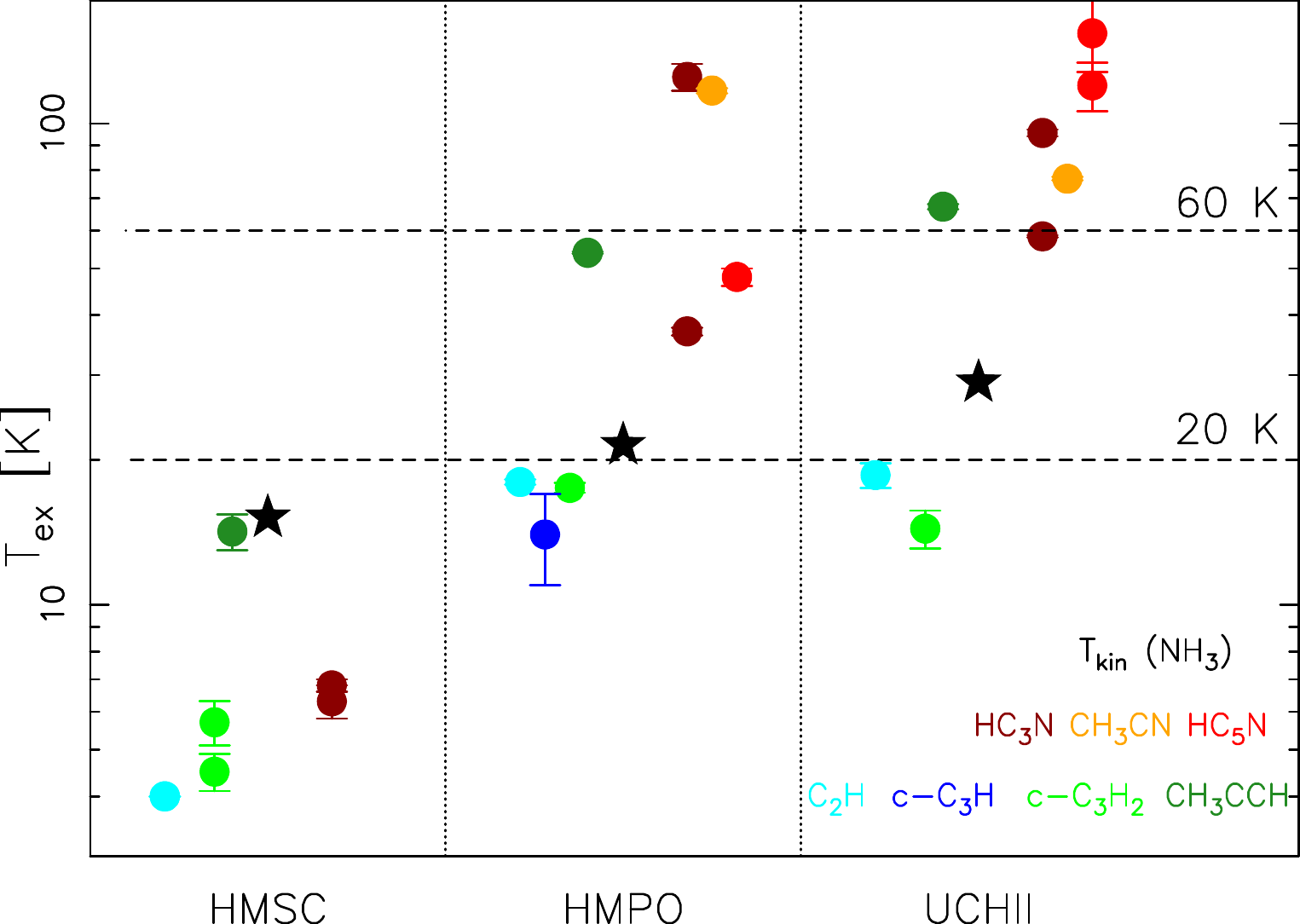}
    \caption{Molecular excitation temperatures. 
    Different colours indicate different species:
    Carbon-chains and hydrocarbons are characterised by blue-cyan-green colours and for each source they are plotted to the left on the x-axis, while molecules with the cyanide group are indicated by red-orange-pink colours and are plotted to the right.
    The black stars are the kinetic temperatures as derived from ammonia.
    For each group, the molecules on the x-axis are ordered from left to right according to increasing number of atoms.}
    \label{fig:tex}
\end{figure}

Figure~\ref{fig:tex} shows all \Tex\ derived in the three targets to highlight trends with the evolutionary stage.
In each source, we show separately hydrocarbons and cyanides, and order the molecules by increasing number of atoms.
The figure also shows \Tk\ in the three targets measured from the inversion transitions (1,1) and (2,2) of ammonia \citep{fontani15a}, which, having critical densities of $\sim 10^{3}$~\cmc\ \citep[e.g.][]{shirley15} are excellent tracers of the kinetic temperature of the moderately dense gas, corresponding to a few $10^3$~\cmc.
There are several trends.
First, in each tracer \Tex\ increases from the HMSC to the later stages, and in some cases (\methyl\ and \CV) also from the HMPO to the UCHII stage.
Second, in each stage the hydrocarbons and carbon-chains tend to be associated with excitation temperatures lower than those of the molecules with the cyanide group, with the exception of \methyl.
As mentioned in Sect.~\ref{afgl}, this is partly due to the fact that hydrocarbons and carbon-chains are observed in transitions with lower $E_{\rm u}$ with respect to cyanides.
However, as highlighted in Table~\ref{tab:energies}, the CHEMICO dataset in principle contains transitions that span a range of $E_{\rm u}$ that goes from $\sim 10$~K up to $\gg 100$~K in all molecules, with the exception of \cIIh, but only the lower excitation ones are detected towards hydrocarbons and small carbon-chains.
Therefore, these species intrinsically trace colder gas better than cyanides.
Third, the higher is the number of atoms in the molecule, the higher \Tex\ tends to be.
In Appendix~\ref{app:fits}, we show a figure similar to Fig.~\ref{fig:tex} plotting the FWHM of the lines.
The increase from the earliest stage to the latest stage is apparent also in FHWM, indicating an increase also in the gas turbulence, whereas there is no obvious trend with the number of atoms in the molecule.

Inspecting the \Tex\ changes in more detail, we see that \HMSC\ shows \Tex\ always smaller than 20~K, and also smaller than (or comparable to in the case of \methyl) the kinetic temperature derived from ammonia ($\sim 15$~K).
\HMPO\ shows a much broader range of \Tex, going from $\sim 14-18$~K in the hydrocarbons \mbox{{\it c-}C$_3$H}, \cyclic, and \cIIh, to $\sim 117-125$~K in \cyanide\ and in the broad component of \CIII\ (see Sect.~\ref{afgl}).
A similar range of \Tex\ is derived in the UCHII \UCHII, but with two significant differences: 
First, the highest temperatures are measured in \CV, while both \cyanide\ and the broad component of \CIII\ both have \Tex $< 100$~K. 
Second, in \UCHII\ an intermediate \Tex\ of $\sim 20-60$~K is missing.
Such trends point to an evolution of the temperature structure.

We propose that an envelope with \Tk\ below or of the order of $\sim 20$~K is present in all evolutionary stages. 
This envelope, traced well by the \AMM\ inversion transitions at low excitation, experiences a marginal temperature increase with evolution, and it is traced preferentially by simple hydrocarbons (\cIIh, \cyclic, \cyclicII).
As the embedded protostar(s) heats up the surrounding gas, an inner protostellar envelope becomes warmer than $20$~K, and the core is now associated with a \Tk\ gradient.
Such envelope is traced by larger molecules and cyanide species, and it is composed itself by a warm (\Tex $\sim 30-60$~K) and a hot (\Tex $\geq 100$~K) component, traced respectively by \methyl, \CV, and the narrow component of \CIII,
and \cyanide\ and the broad component of \CIII.
Finally in the UCHII phase, \Tex\ of the cyanide species and of the complex hydrocarbon \methyl\ increases further.
In this late phase, only gas with \Tex\ above 60~K and below 20~K is present.
Such temperature steepening could be due to the fact that the envelope has had more time to warm up at larger distances from the core centre.
The only two species showing a countertrend, that is a \Tex\ decrease between the HMPO and the UCHII stage, are \cyanide\ and the broad component of \CIII.
We speculate that this countertrend is due to the fact that in the UCHII phase the innermost part of the hot-core, which is the hottest portion of the molecular environment of the nascent protostar(s), is destroyed by the ionised expanding region.
This reduces the hot molecular cocoon heated during the HMPO stage to a thinner region in the UCHII stage.
Because this thinner hot-core is further away from the central nascent star(s), the average \Tex\ is lower than that of the intact core in the HMPO stage.
The innermost ionised region of \UCHII\ is traced in the CHEMICO spectra by Hydrogen recombination lines. 
We detected from the H44$\alpha$ line at $\sim 74.644$~GHz to the H29$\alpha$ line at $\sim 256.302$~GHz.
The FWHM ($\sim 60$~\kms) of these lines is consistent with a very hot gas, confirmed by a preliminary analysis of the excitation temperature of these lines, which exceeds 2000~K.
We will show the analysis of these recombination lines in a forthcoming paper (Fontani et al. in prep.).

A sketch of the temperature structure of \UCHII\ based on the CHEMICO data, and the comparison with the earlier evolutionary stages, is shown in Fig.~\ref{fig:AFGL}.

\subsection{Evolution of the fractional abundances}
\label{discu-abb}

\begin{figure}
    \centering
    \includegraphics[width=1\linewidth]{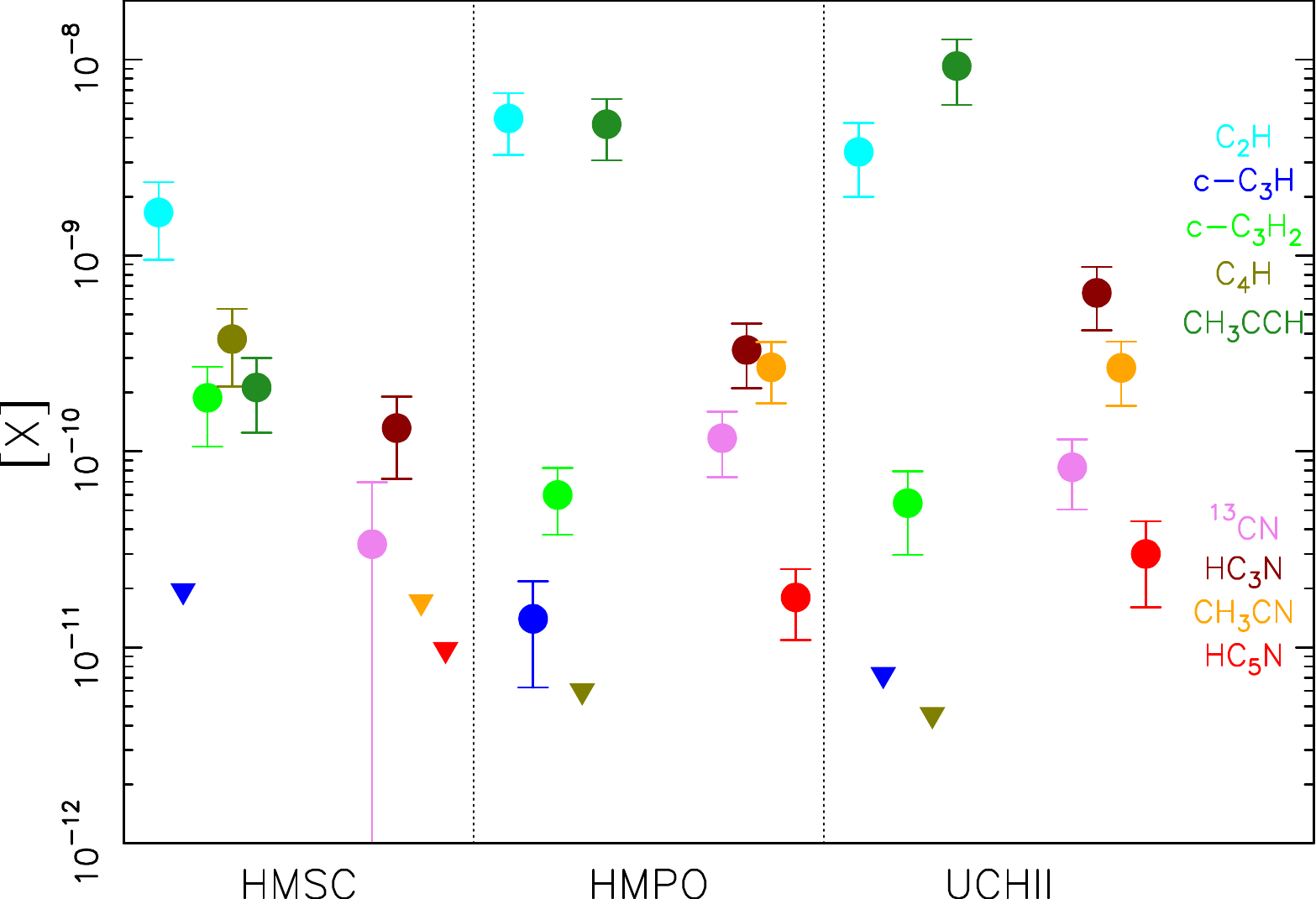}
    \caption{Molecular fractional abundances. 
    As in Fig.~\ref{fig:tex}, hydrocarbons are characterised by blue-cyan-green colours and in each source are grouped to the left on the x-axis, molecules with the cyanide group are indicated by red-orange-pink colours, and the molecules on the x-axis are ordered from left to right according to increasing number of atoms (or carbon atoms in case of equal total number of atoms).
    Upper limits are indicated by triangles pointing downwards.}
    \label{fig:abb}
\end{figure}

Figure~\ref{fig:abb} shows all fractional abundances in the three targets, including upper limits for the non-detections.
For species in which two velocity components were detected, the abundances were computed adding the column densities of the two components.
As in Fig.~\ref{fig:tex}, in each source we separately show hydrocarbons and cyanides, and order the molecules by their number of atoms.
As reference for cyanide species, we also plot the fractional abundance of $^{13}$CN.
Overall, $X$ of the simple species do not change significantly in the three sources.
The only simple molecule associated with a significant increase from the HMSC phase to the later stages is \cIIh.
However, the increase is less than a factor $\sim 4$ and the \Tex\ used to derive \Ntot\ in \HMSC\ had to be fixed, as explained in Sect.~\ref{g034}.
The only species that show a significant abundance enhancement with evolution are the two COMs, \methyl\ and \cyanide.
In particular, $X$ increases by more than an order of magnitude from the HMSC to the HMPO stage.
Such increase is likely associated with the evaporation of ice mantles from dust grains that releases COMs and their precursors in the gas phase.

\begin{figure}
    \centering
    \includegraphics[width=1\linewidth]{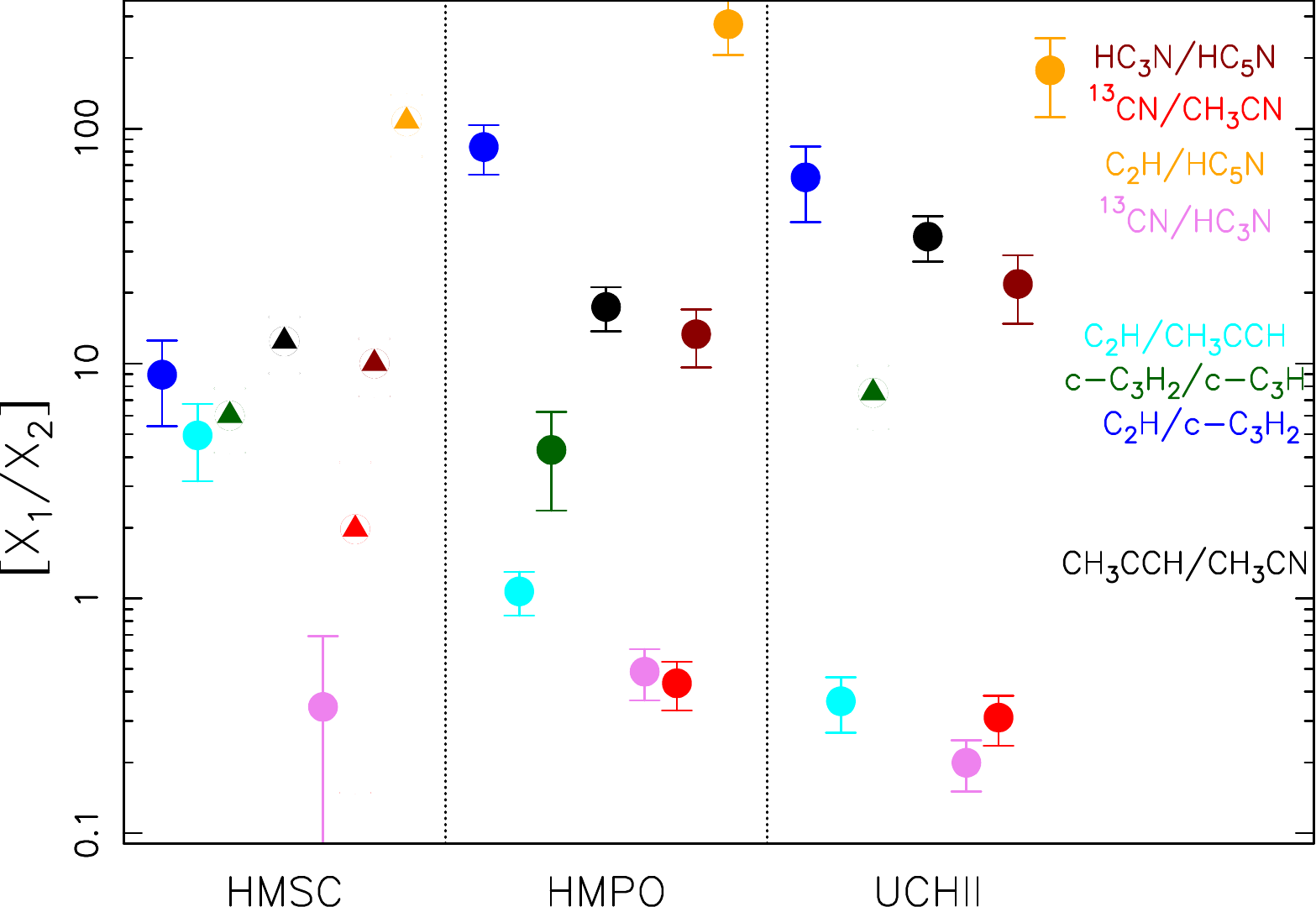}
    \caption{Molecular abundance ratios of chemically related species. 
    As in Fig.~\ref{fig:tex}, hydrocarbons are characterised by blue-cyan-green colours and in each source are grouped to the left on the x-axis, molecules with the cyanide group are indicated by red-orange-pink colours.
    The ratio between the two COMs \methyl\ and \cyanide\ are indicated in black.
    Lower limits are indicated by triangles pointing upwards.}
    \label{fig:ratios}
\end{figure}

Some column density ratios of species that are tought to be chemically related (e.g. parent/daughter species, or species formed similarly) are shown in Fig.~\ref{fig:ratios}. 
To simplify the plot, we computed only the ratios related to the strongest velocity feature in \HMSC, and the narrow velocity features in \HMPO\ and \UCHII.
The ratio that shows a clear decrease with evolution is only C$_2$H/CH$_3$CCH, with C$_2$H being more abundant than \methyl\ in the HMSC, and becoming less abundant in the UCHII.
C$_2$H is a gas-phase precursor of \methyl, thus a decrease could indicate a progressive destruction of \cIIh\ to form \methyl.
A tentative decrease is also seen in $^{13}$CN/\cyanide, and as CN is also a progenitor of \cyanide\ the tentative decrease can be due to the same reason.
Among the ratios that show a significant increase is the ratio between the two COMs, the \methyl/\cyanide\ ratio is enhanced by a factor $\sim 2.5$ from the HMPO to the UCHII stage. 
The \CIII/\CV\ ratio is also increasing by a factor of 1.7, but it is within the error bars.
Finally, the ratios that show an increase from the HMSC phase to the HMPO phase, and then a decrease to the UHCII phase, are \cIIh/\cyclic, $^{13}$CN/\CIII, $^{13}$CN/\CIII, and $^{13}$CN/\cyanide.
In particular, the \cIIh/\cyclic\ ratio increases by an order of magnitude from the HMSC to the HMPO phase.
This increase is only due to the increase of the \cIIh\ abundance, since the abundance of \cyclic\ remains constant (Fig.~\ref{fig:abb}).

\subsection{Comparison with other evolution studies}
\label{discu-comp}

As discussed in \citet{taniguchi24}, carbon-chain molecules are claimed to be evolutionary indicators in low-mass star-forming cores, but it is still unclear if this is true also in the high-mass case.
Based on chemical simulations, \citet{taniguchi21} have proposed that the \cIIh/\CV\ ratio decreases as the temperature increases.
In our data, such a decrease is seen between the HMPO and the UCHII stage, but not between the HMSC and the HMPO stage (Fig.~\ref{fig:ratios}).
Moreover, the derived ratios in between 100 and 300 are in between what was measured towards the low-mass star-forming core L1527 ($\sim 600$) and the high-mass cores observed by \citet{taniguchi21} ($\sim 15$).
This indicates a large variation of this ratio from one source to another, making it difficult to claim it as an evolutionary indicator.

The ratio between the two cyanopolyynes, \CIII/\CV, can also potentially be an indicator of evolution.
In particular, \citet{fontani17} have found that such a ratio is sensitive to variations of the cosmic-ray ionisation rate, $\zeta$: the higher $\zeta$ is, the lower the \CIII/\CV\ should be.
Because protostars are sources of local cosmic-rays \citep[e.g.][]{padovani16}, $\zeta$ close to protostellar objects is believed to be enhanced with respect to that associated with quiescent phases, and therefore the \CIII/\CV\ ratio should decrease from the HMSC phase to the later phases.
Although we did not detect \CV\ in \HMSC, the ratio does not show a decrease with evolution, perhaps indicating that the effect of local cosmic rays is not effective on the abundances of cyanopolyynes on the large angular scales of our observations.

\citet{mininni21} found that the abundance of \cyanide\ is an evolutionary indicator for high-mass star-forming cores, since its abundance jumps by more than an order of magnitude between the HMSC and the HMPO stage.
We confirm this finding and also the upper limit for the abundance of \cyanide\ in high-mass starless cores, which is $\leq 4\times 10^{-11}$, higher than what we measure towards \HMSC.

\section{Conclusions and outlook}
\label{conc}

We have presented an overview of the project "CHemical Evolution of MassIve star-froming COres" in which,
thanks to an unbiased full spectral survey of the 3, 2, and 1.2~mm receiver bands of the IRAM 30m telescope, we investigate any aspect of the chemical evolution of three dense cores in different evolutionary stages. 
The chemical richness is found to increase with evolution in terms of both species and lines detected.
In this first paper, we analyse the carbon rich species \cIIh, \cyclicII, \cyclic, \cIVh, \methyl, \CIII,  \cyanide\ and \CV, and derive from them the temperature structure variation using a LTE approach. 
The hydrocarbons and carbon-chains are associated with \Tex\ lower than those of cyanides. 
Moreover, the higher the number of atoms in the molecule, the higher \Tex.
The fractional abundances with respect to H$_2$ show significant enhancements with evolution only for the two COMs \methyl\ and \cyanide. 

We propose that the temperature structure evolves with time as follows: the HMSC stage is characterised by a cold ($\sim 20$~K) uniform temperature envelope, preferentially traced by simple hydrocarbons (\cIIh, \cyclic, \cyclicII).
The protostellar heating in the HMPO stage creates a temperature structure which can be outlined as an innermost hot core (\Tex $\geq 100$~K) traced by \cyanide\ and the broad components of cyanopolyynes, surrounded by a shell with intermediate temperature ($30-60$~K) traced by \methyl, \CV\ and the narrower \CIII\ component, and a cold extended envelope unaffected by protostellar heating as in the HMSC stage.
Finally in the UCHII phase, \Tex\ of the cyanide species and of the complex hydrocarbon \methyl\ increase further, and the inner layer with intermediate temperature disappears.

This first work shows the power of CHEMICO in determining the change in not only the chemical complexity but also the physical structure of the sources with evolution.
Follow-up high(er)-angular resolution studies will be essential to better investigate the core-scale chemical and physical structure of the three CHEMICO targets, needed for a proper comparison with chemical models.

\begin{acknowledgements}

This work is based on observations carried out under project number [124-20] and [129-21] with the IRAM 30m telescope. 
IRAM is supported by INSU/CNRS (France), MPG (Germany) and IGN (Spain).
We acknowledge the IRAM staff for help provided during the observations.
V.M.R. and L.C. acknowledge support from the grant PID2022-136814NB-I00 by the Spanish
Ministry of Science, Innovation and Universities/State Agency of Research MICIU/AEI/10.13039/501100011033 and by ERDF, UE.
V.M.R. also acknowledges support from the grant RYC2020-029387-I funded by MICIU/AEI/10.13039/501100011033 and by "ESF, Investing in your future", and from the Consejo Superior de Investigaciones Cient{\'i}ficas (CSIC) and the Centro de Astrobiolog{\'i}a (CAB) through the project 20225AT015 (Proyectos intramurales especiales del CSIC); and from the grant CNS2023-144464 funded by MICIU/AEI/10.13039/501100011033 and by “European Union NextGenerationEU/PRTR”. 
H.M-C. acknowledges support from the grant JAE-intro (JAEINT-23-01443) from the Consejo Superior de Investigaciones Científicas (CSIC).
The research leading to these results has received funding from the European Union’s Horizon 2020 research and innovation program under grant agreement No 101004719.

\end{acknowledgements}

%
%

\onecolumn

\begin{appendix} 
\FloatBarrier
\section{Fit results for $V_{\rm p}$ and FWHM}
\label{app:fits}

\begin{figure}[h!]
    \centering
    \includegraphics[width=0.65\linewidth]{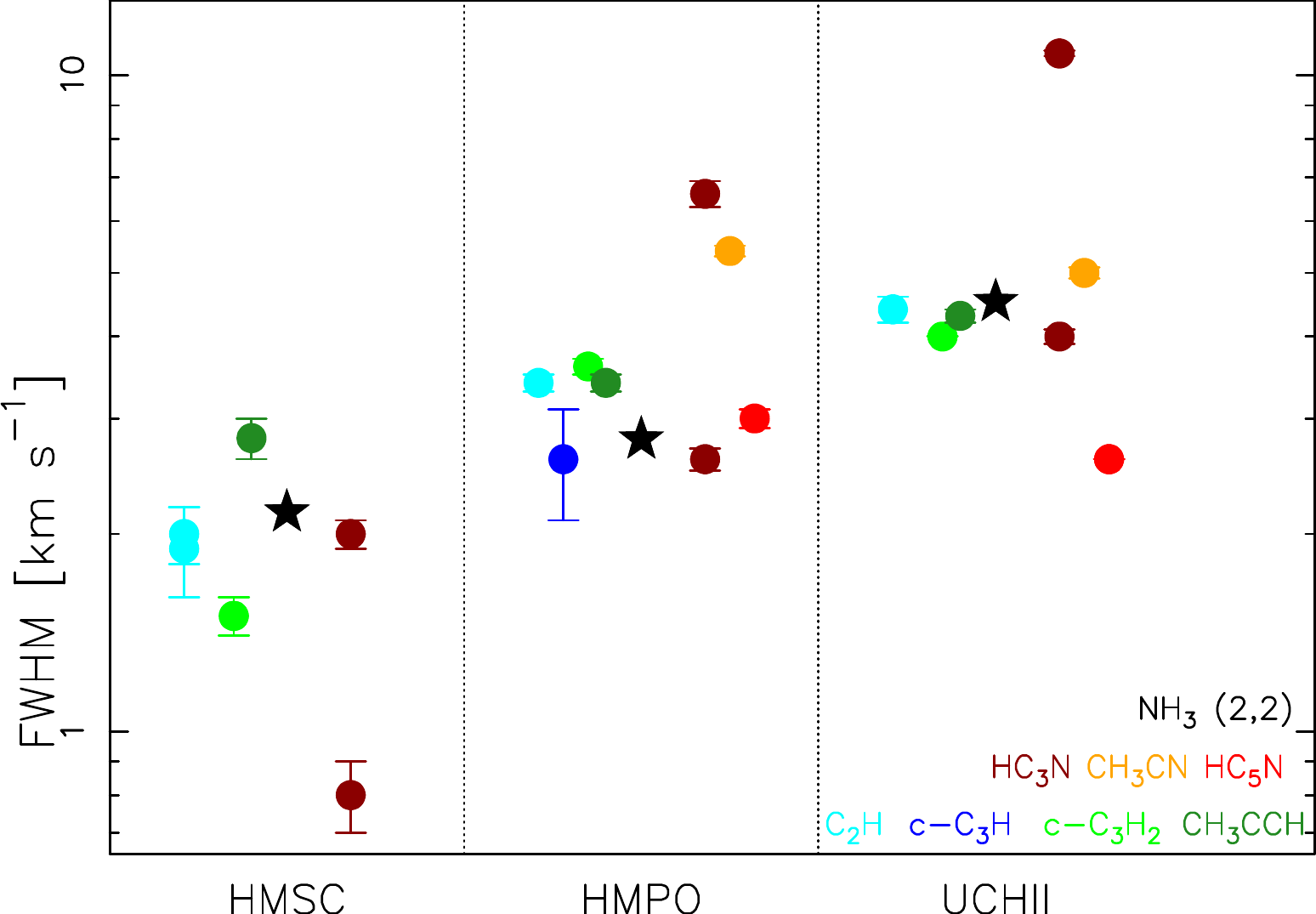}
    \caption{Line FWHM of all molecules detected in each tracer.
    Different colours indicate different species as in Fig.~\ref{fig:tex}:
    carbon-chains and hydrocarbons are characterised by blue-cyan-green colours and for each source they are plotted to the left on the x-axis; molecules with the cyanide group are indicated by red-orange-pink colours and are plotted to the right.
    The black stars are the FWHM derived from ammonia (2,2) \citep{fontani15a}.
    For each group, the molecules on the x-axis are ordered from left to right according to increasing number of atoms.}
    \label{fig:Dv}
\end{figure}

Best fit peak velocity and full width at half maximum obtained with {\sc madcuba} for the lines analysed in this paper.
In Figure~\ref{fig:Dv}, we also plot the FWHM of the lines derived for all molecules in each source.
The molecules are plotted as in Fig.~\ref{fig:tex}.

\FloatBarrier
\begin{table}[h!]
\begin{center}
\setlength{\tabcolsep}{2.5pt}
\caption{\label{tab:fit-v} Best-fit $V_{\rm p}$.}
\begin{tabular}{l l l l l l l l l}
\hline
\hline
  Source  & \multicolumn{8}{c}{$V_{\rm p}$ (\kms)} \\
      & \cIIh & \cyclicII & \cyclic & C$_4$H &  \methyl & \CIII\ & \CV\  & \cyanide \\
\hline
G034-1 & 41.3\tablefootmark{a} & -- & 41.3(0.1) & 41.3\tablefootmark{a} & 42.3(0.1)\tablefootmark{b} & 41.7(0.1) & -- & -- \\
G034-2 & 43.3\tablefootmark{a} & -- & 43.3(0.1) & 43.3\tablefootmark{a} & -- & 43.4(0.1) & -- & -- \\
AFGL-1 & --2.6(0.1) & --2.9(0.2) & --2.7(0.1) & --3.1(0.3) & --2.7(0.1) & --2.7(0.1) & --2.6(0.1) & --2.7(0.1) \\
AFGL-2 & -- & -- & -- & -- & -- & --2.7(0.1) & -- & -- \\
G589-1 & 9.0(0.1) & -- & 8.8\tablefootmark{a} & -- & 8.9(0.1) & 8.8(0.1) & 7.8(0.2) & 9.3(0.1) \\
G589-2 & -- & -- & -- & -- & -- & 11.5(0.1) & 11(0.3) & -- \\
\hline
\end{tabular}
\end{center}
\tablefoot{
\tablefoottext{a}{Fixed;}
\tablefoottext{b}{The two velocity features are spectrally unresolved.}
}
\end{table}

\FloatBarrier
\begin{table}[h!]
\begin{center}
\setlength{\tabcolsep}{2.5pt}
\caption{\label{tab:fit-fwhm} Best-fit FWHM.}
\begin{tabular}{l l l l l l l l l}
\hline
\hline
  Source  & \multicolumn{8}{c}{FWHM (\kms)} \\
      & \cIIh & \cyclicII & \cyclic & C$_4$H &  \methyl & \CIII\ & \CV\  & \cyanide \\
\hline
G034-1 & 2.0(0.2) & -- & 1.5(0.1) & 1.5\tablefootmark{a} & 2.8(0.2)\tablefootmark{b} & 2.0(0.1) & -- & -- \\
G034-2 & 1.9(0.3) & -- & 1.5(0.1) & 1.5\tablefootmark{a} & -- & 0.8(0.1) & -- & -- \\
AFGL-1 & 3.4(0.1) & 2.6(0.5) & 3.6(0.1) & 4.0\tablefootmark{a} & 3.4(0.1) & 2.6(0.1) & 3.0(0.1) & 5.4(0.1) \\
AFGL-2 & -- & -- & -- & -- & -- & 6.6(0.3) & -- & -- \\
G589-1 & 4.4(0.2) & -- & 4.0\tablefootmark{a} & -- & 4.3(0.1) & 4.0(0.1) & 2.6\tablefootmark{a} & 5.0(0.1) \\
G589-2 & -- & -- & -- & -- & -- & 10.8(0.1) & 2.6\tablefootmark{a} & -- \\
\hline
\end{tabular}
\end{center}
\tablefoot{
\tablefoottext{a}{Fixed;}
\tablefoottext{b}{The two velocity features are spectrally unresolved.}
}
\end{table}

\FloatBarrier

\FloatBarrier
\section{Spectra of representative lines}
\label{app:spectra}
\FloatBarrier

Sample spectra of lines representatives of the species analysed in this paper (except \CIII, whose lines are shown in Figs.~\ref{fig:HC3N-G034},~\ref{fig:spectra-HMPO}, and~\ref{fig:HC3N-G589}) for each source.
\label{app:spectra}

\FloatBarrier
\begin{figure}[h!]
    \centering
    \includegraphics[width=0.75\linewidth]{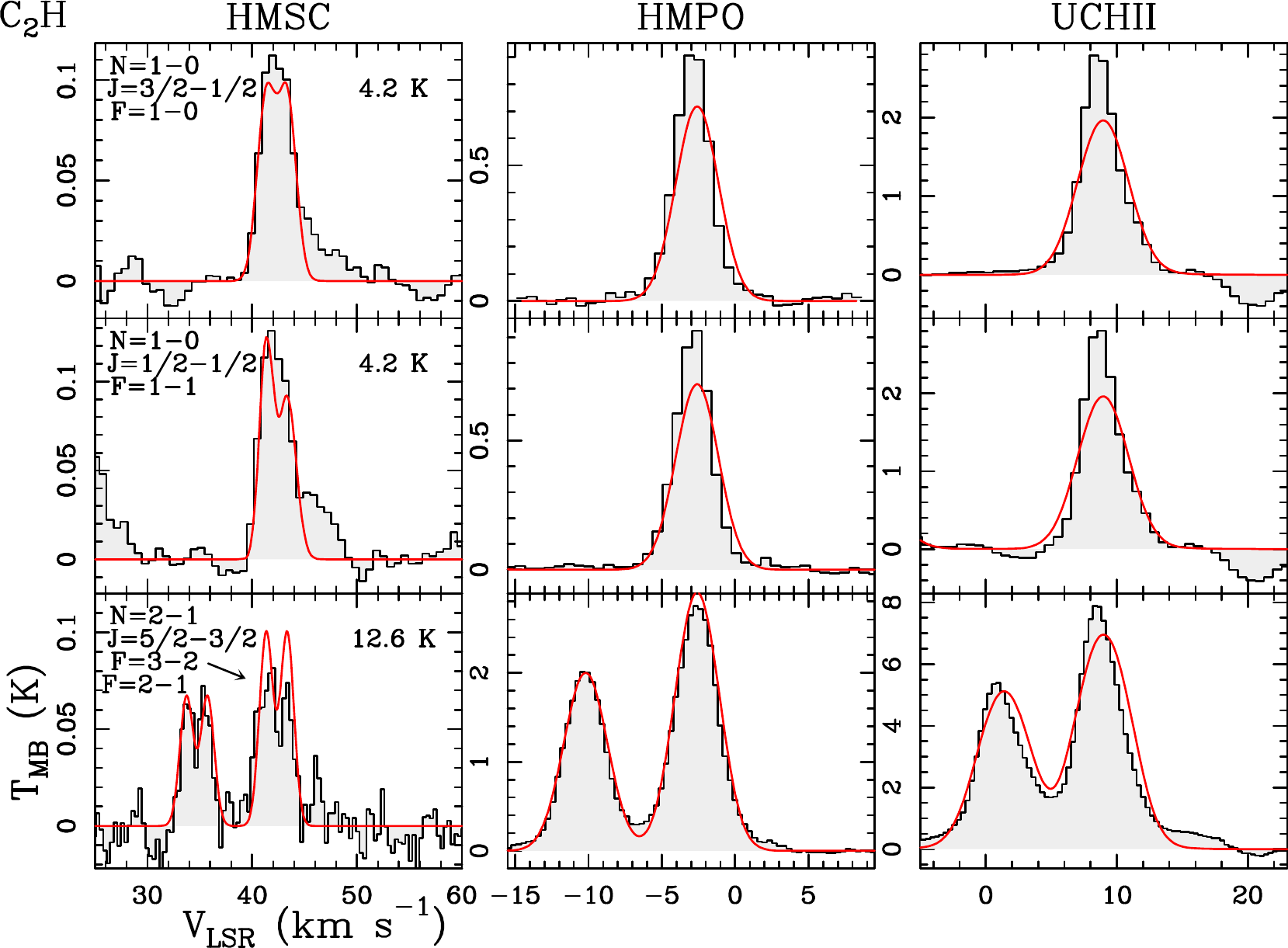}
    \caption{Spectra of a sample of \cIIh\ lines. The red curves are the best LTE fits obtained with {\sc madcuba}. Towards \HMSC, the total fit obtained towards the two velocity components is shown.
    Quantum numbers and energy of the upper level of the transition are shown in the top left and top right corners, respectively, of the HMSC line plots.}
    \label{fig:C2H-spectra}
\end{figure}

\FloatBarrier
\begin{figure}[h!]
    \centering
    \includegraphics[width=0.75\linewidth]{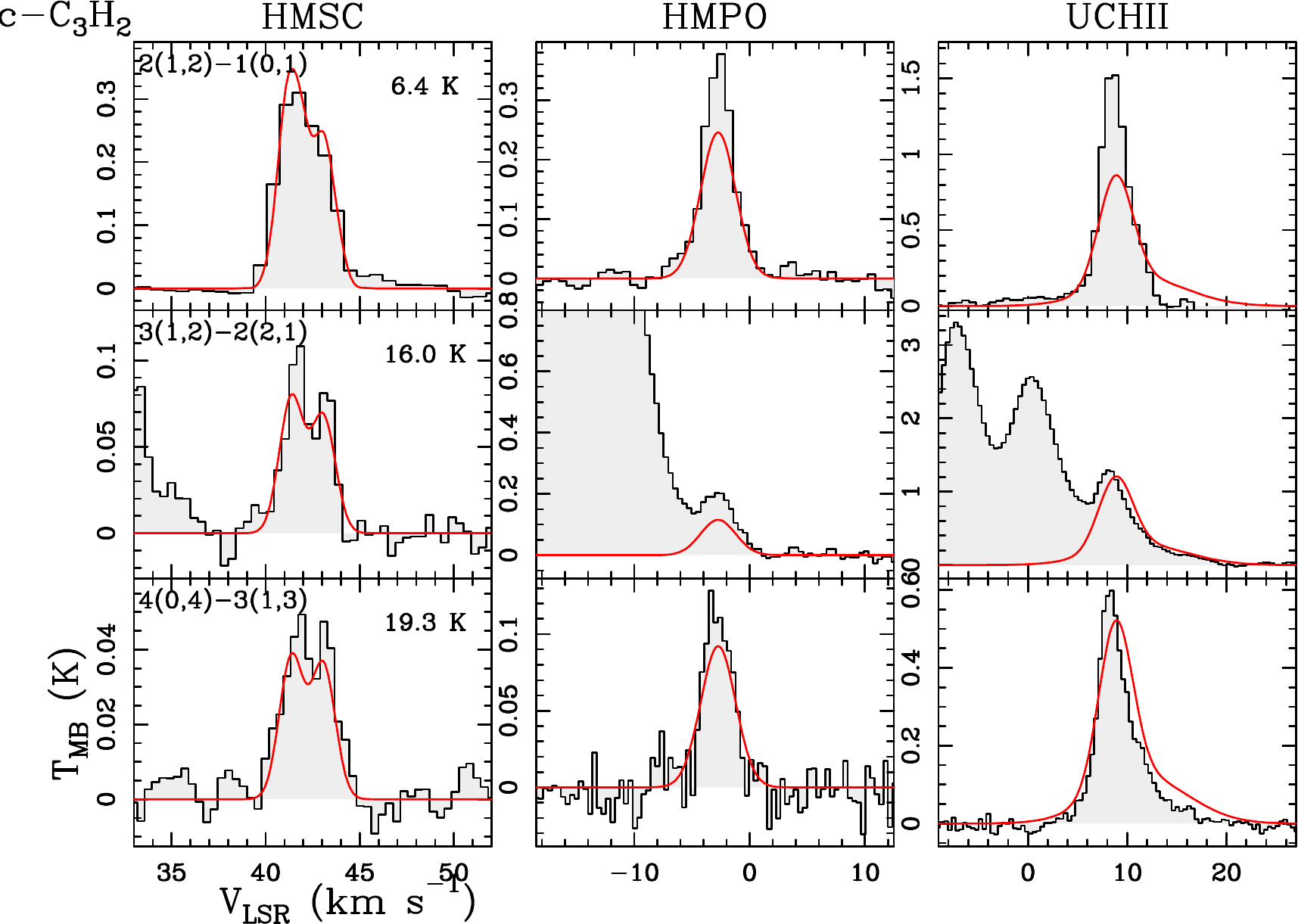}
    \caption{Same as Fig.~\ref{fig:C2H-spectra} for \cyclic.}
    \label{fig:C3H2-spectra}
\end{figure}

\FloatBarrier
\begin{figure}[h!]
    \centering
    \includegraphics[width=0.75\linewidth]{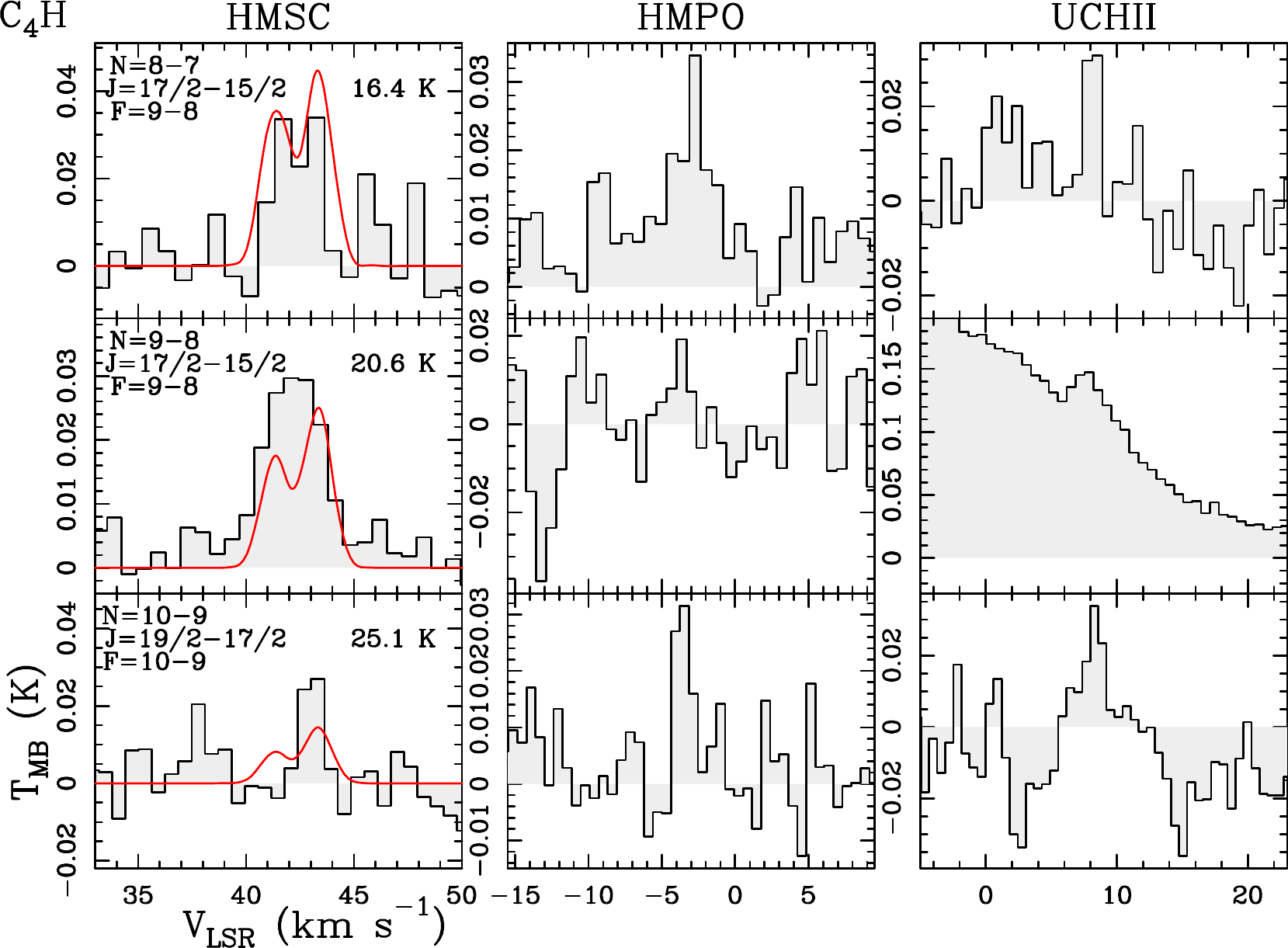}
    \caption{Same as Fig.~\ref{fig:C2H-spectra} for \cIVh.}
    \label{fig:C4H-spectra}
\end{figure}

\FloatBarrier
\begin{figure}[h!]
    \centering
    \includegraphics[width=0.75\linewidth]{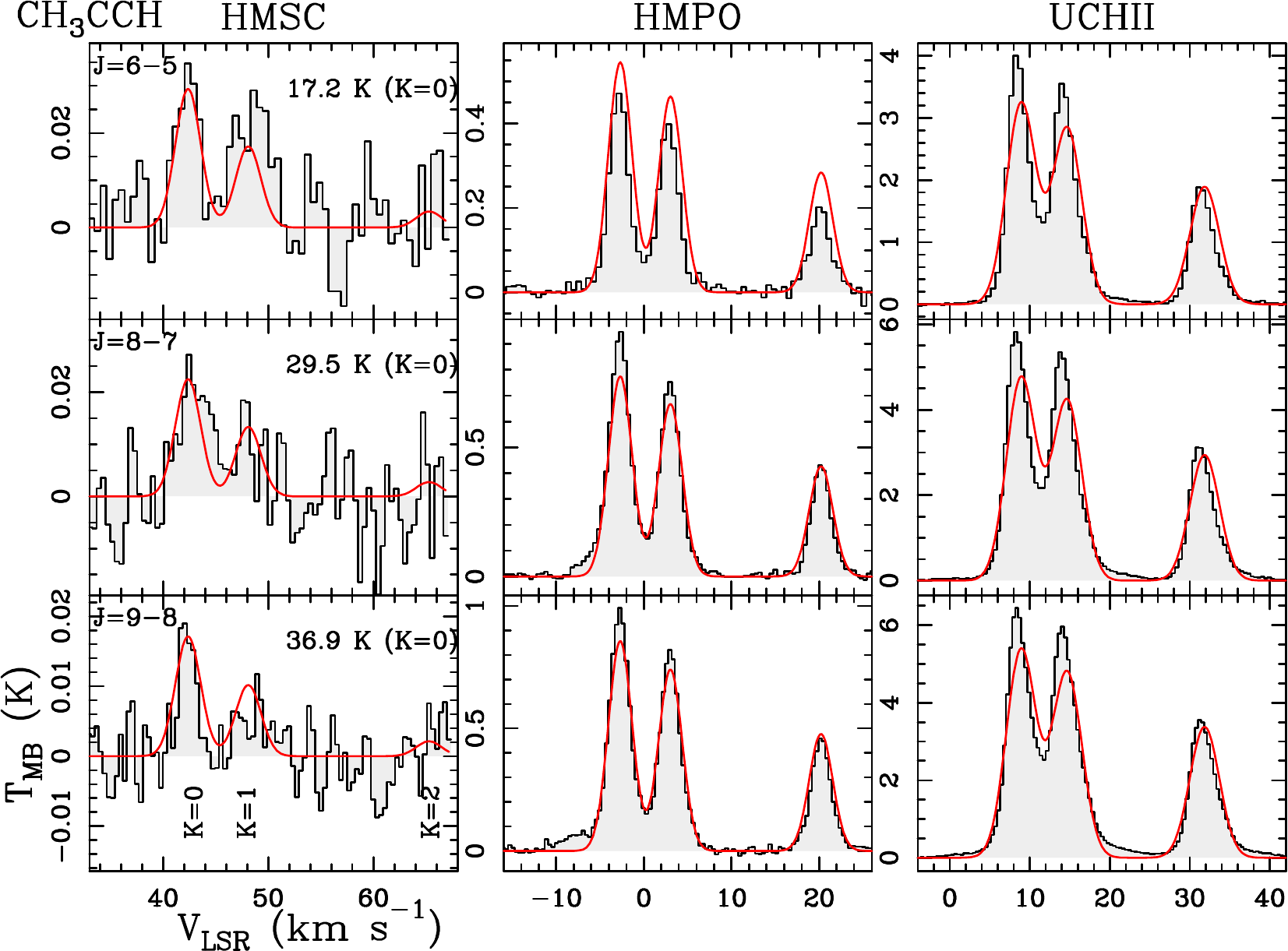}
    \caption{Same as Fig.~\ref{fig:C2H-spectra} for \methyl.}
    \label{fig:CH3CCH-spectra}
\end{figure}

\FloatBarrier
\begin{figure}[h!]
    \centering
    \includegraphics[width=0.75\linewidth]{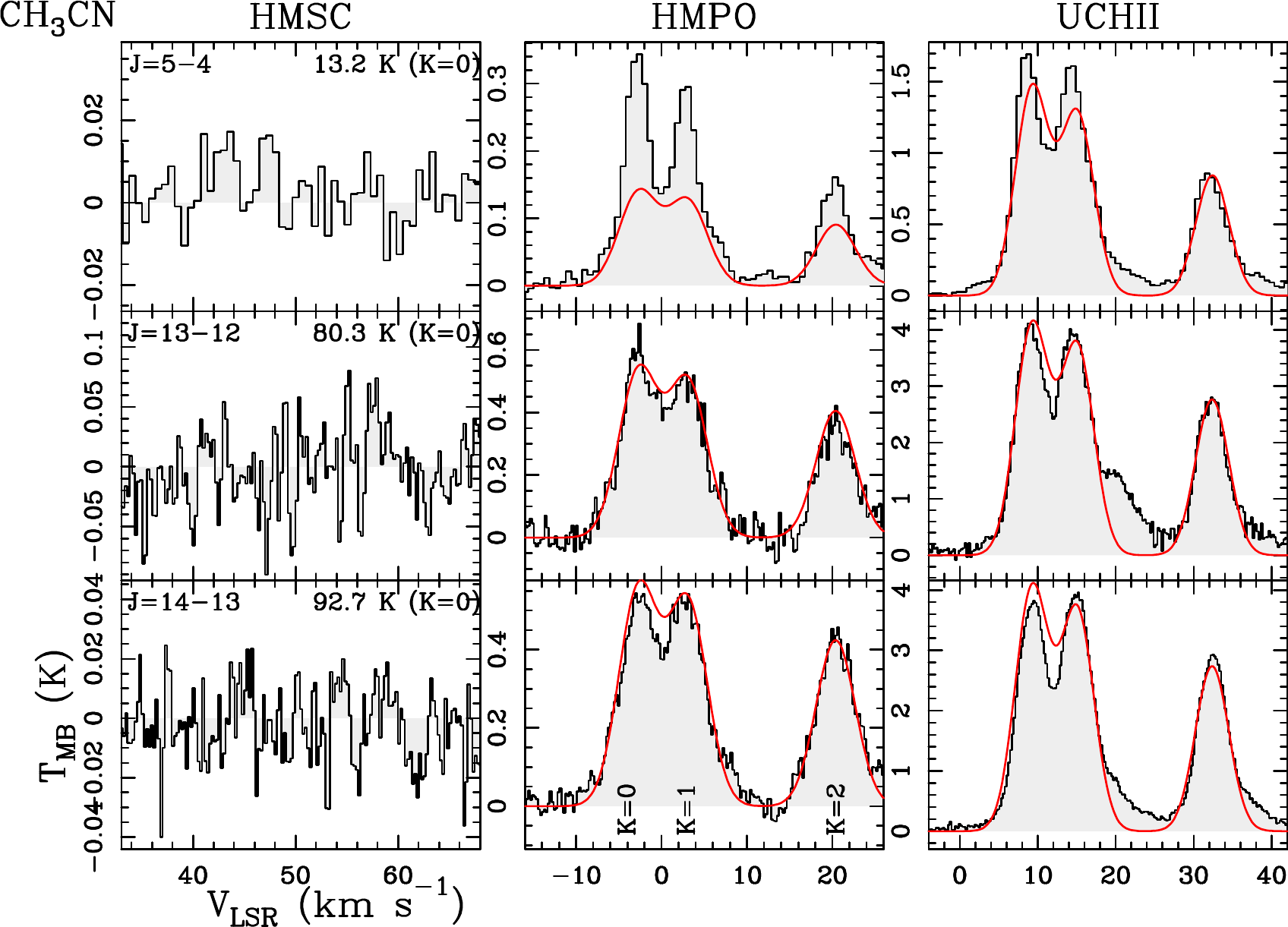}
    \caption{Same as Fig.~\ref{fig:C2H-spectra} for \cyanide.}
    \label{fig:CH3CN-spectra}
\end{figure}

\FloatBarrier
\begin{figure}[h!]
    \centering
    \includegraphics[width=0.75\linewidth]{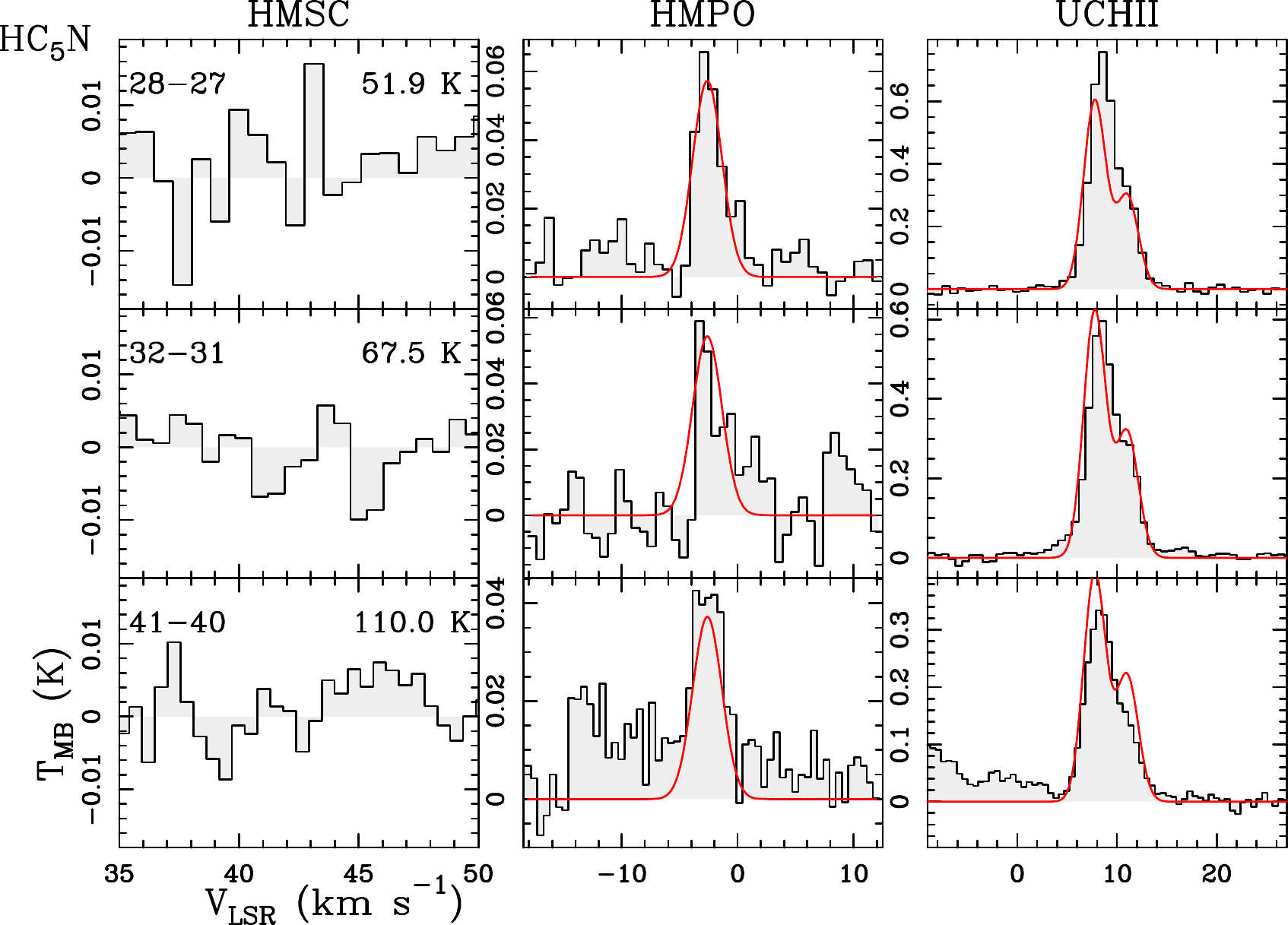}
    \caption{Same as Fig.~\ref{fig:C2H-spectra} for \CV.}
    \label{fig:HC5N-spectra}
\end{figure}

\end{appendix}

\end{document}